\documentclass[useAMS,graphics,usenatbib]{mn2e}

\usepackage{amssymb,color}
\usepackage{epsfig}
\usepackage{times} 

\newcommand{\galform}{\textsc{GALFORM}}

\newcommand{\Baugh}{\textsc{Baugh05}}
\newcommand{\Bower}{\textsc{Bower06}}
\newcommand{\mill}{\textsc{Millennium}}

\newcommand{\dd}{{\rm d}}
\newcommand{\refeq}[1]{(\ref{#1})}  
\newcommand{\mr}[1]{{\rm #1}}     

\newcommand{\msun}{h^{-1}{\rm M_{\odot}}}

\title[The galaxies that reionized the Universe]{The galaxies that reionized the Universe}
\author[Rai\v{c}evi\'{c} et al.]{Milan Rai\v{c}evi\'{c}$^{1,2}$\thanks{E-mail:
milan.raicevic@durham.ac.uk}, Tom Theuns$^{1,3}$, Cedric Lacey$^1$\\
$^1$Institute for Computational Cosmology, Durham University, Science Laboratories, Durham DH1 3LE,
UK\\
$^2$Leiden Observatory, Leiden University, P.O. Box 9513, 2300RA Leiden, The Netherlands\\
$^3$Universiteit Antwerpen, Campus Groenenborger, Groenenborgerlaan
171, B-2020 Antwerpen, Belgium\\ 
}

\begin{document}

\date{}

\pagerange{\pageref{firstpage}--\pageref{lastpage}} \pubyear{2009}

\maketitle

\label{firstpage}

 \begin{abstract}
   The Durham \galform\ semi-analytical galaxy formation model has
   been shown to reproduce the observed rest-frame 1500\AA\
   luminosity function of galaxies well over the whole redshift range
   $z=5-10$.  We show that in this model, this galaxy population also
   emits enough ionizing photons to reionize the Universe by redshift
   $z=10$, assuming a modest escape fraction of 20 per cent. The bulk
   of the ionizing photons is produced in faint galaxies during
   starbursts triggered by galaxy mergers. The bursts introduce a
   dispersion up to $\sim 5$~dex in galaxy ionizing luminosity at a
   given halo mass. Almost 90 per cent of the ionizing photons emitted
   at $z=10$ are from galaxies below the current observational
   detection limit at that redshift. Photo-ionization suppression of
   star formation in these galaxies is unlikely to affect this
   conclusion significantly, because the gas that fuels the starbursts
   has already cooled out of their host halos. The galaxies that
   dominate the ionizing emissivity at $z=10$ are faint, with
   $M_{1500, {\rm AB}}\sim -16$, have low star formation rates, $\dot M_\star\sim 0.06\,h^{-1}M_\odot$~yr$^{-1}$, and reside in halos of mass $M\sim 10^9\,h^{-1}M_\odot$.
\end{abstract}


\section{Introduction}
Reionization refers to the transition in the state of the Universe
from mostly neutral, following recombination at a redshift of $z\sim
1000$, to highly ionized once more at later times.
\citet{GunnPeterson} \citep[and also][]{BahcallSalpeter} realised as soon as \citet{Schmidt} published spectra of $z\sim 2$ quasars that the absence of significant Lyman-$\alpha$ absorption in their spectra
implied that the $z\sim 2$ Universe is very highly ionized.  That
basic picture has not changed with the discovery by \citet{Fan03} of
$z>6$ QSOs \citep{Becker07} or the novel method based on gamma-ray
bursts as probes of the intergalactic medium (IGM) at even higher $z$
\citep{Totani06,Zafar10,Patel10}.

The fact that most of the hydrogen in the Universe is highly ionized at
least as early as $z\sim 7$ is also consistent with the large Thomson
scattering optical depth toward the surface of large scattering which
is inferred from measurements of CMB fluctuations. This implies a \lq
reionization redshift\rq\ of $z_{\rm reion}=10.5\pm 1.2$, if the
transition from neutral to completely ionized occurred instantaneously
\citep{Komatsu10}. The temperature of the IGM depends on its
reionization history because the thermal timescales are long:
measurements of that temperature \citep{Schaye00} are also consistent
with reionization occurring around $z\sim 10$ \citep{Theuns02}.

The current paradigm as to how reionization happens is that initially
small H{\sc II} regions form around individual sources of ionizing
photons\footnote{However, a strong background flux of higher-energy
radiation, for example X-rays from accreting black holes, may \lq
pre-reionize\rq\ the Universe \citep{Oh01}.}. As the sources become
brighter and more numerous, isolated H{\sc II} regions grow, merge,
and eventually percolate throughout the IGM, see for example the early
simulations by \citet{Gnedin97}. The nature of the sources of
the ionizing radiation is still unknown. While a number of works show
that the majority of ionizing radiation is probably produced by
stellar sources \citep[e.g.][]{madau99, Gnedin00a, Sokasian03,
Ciardi03, Furlanetto04, Trac07, Trac09}, the exact contribution of
Population III stars or quasars is under debate \citep[see][for recent
examples]{Choudhury07, Wyithe07, loeb09, Volonteri09, Salvaterra10}.

Depending on its spin temperature, the not-yet ionized H{\sc I} during
the epoch of reionization (EoR) could be detected in either emission
or absorption in redshifted 21-cm radiation, either in the form of a
global step in the spectrum, or indeed probing the remaining neutral
regions in a partly ionized IGM \citep{Madau97,Shaver99,Tozzi00}.
Because most plausible ionizing sources will be highly clustered, the
ionized bubbles could grow to be quite large, and the epoch where the
IGM is 50\% ionized may be best suited for direct detection with
current and future experiments, such as
LOFAR\footnote{http://www.lofar.org/},
21CMA\footnote{http://21cma.bao.ac.cn/},
MWA\footnote{http://www.haystack.mit.edu/ast/arrays/mwa/}, and
eventually the SKA\footnote{http://www.skatelescope.org/}.  The
promise of a direct observational probe has stimulated considerable
interest in the EoR, see recent reviews by for example
\citet{Barkana01, Ciardi05, Loeb06, Trac09}.

The recent installation of the Wide Field Camera 3 (WFC3)/IR on the
{\em Hubble Space Telescope} has made it possible to search for $z>6$
galaxies using the \lq Lyman-break\rq\ drop-out technique, with a
number of authors reporting the discovery of galaxies with $z>6$
(based on their colours), with candidates up to $z\sim 10$
\citep{Bunker09a, Bouwens07, Bouwens09a, Bouwens09c, Oesch09}. Are
these the galaxies that caused reionization? The analysis by
\citet{Bunker09b} suggests that these galaxies are unlikely to produce
sufficient ionizing photons to reionize the Universe. In fact even at
lower $z\sim 6$ there seems to be a problem, in the sense that the
observed galaxies do not appear to produce sufficient photons to keep
the IGM from recombining \citep{Bolton07}.

In this paper we use the \galform\ semi-analytical model of galaxy
formation \citep{Cole00, Baugh05} to make theoretical predictions for
the evolution of the emissivity of ionizing photons from galaxies,
$\epsilon(z)$. The \galform\ model calculates the formation and
evolution of galaxies in the framework of hierarchical structure
formation in CDM, including baryonic physics such as gas cooling, star
formation and supernova feedback.  In contrast to most other
work modelling the contribution of galaxies to reionization, the
\galform\ model which we use here was originally developed to try to
explain the properties of galaxies at much lower
redshifts. Predictions from \galform\ have been compared with a very
wide range of observational data at lower redshifts $z \lesssim 6$,
including galaxy luminosity functions, colours, sizes, morphologies,
gas contents and metallicities at redshift $z=0$
\citep{Cole00,Baugh05,Bower06,Gonzalez09}, the evolution of galaxies
at optical, IR and sub-mm wavelengths
\citep{Baugh05,Bower06,Lacey08,Gonzalez-Perez09}, and the evolution of
$Ly\alpha$-emitting galaxies \citep{LeDelliou06,Orsi08}. In this
paper, we use the \citet{Baugh05} variant of the \galform\ model. This
model was already shown by \citeauthor{Baugh05} to reproduce the
observed rest-frame far-UV luminosity function of Lyman-break galaxies
at $z\sim 3$, and the same model has recently been shown to reproduce
the observed numbers of Lyman-break galaxies over the whole range $z
\sim 3-10$ \citep{Lacey10b}. Two important features of the
\citet{Baugh05} model are that at high redshifts, most star formation
happens in starbursts triggered by galaxy mergers, and the initial
mass function (IMF) of the stars formed in such bursts is top heavy,
containing a much larger proportion of high-mass stars than is found
in more quiescent star formation environments such as the solar
neighbourhood. This top-heavy IMF was introduced into the model by
\citeauthor{Baugh05} in order to explain the observed numbers and
redshifts of the faint sub-mm galaxies, now known to be very luminous,
dust-obscured starbursts at $z \sim 1-3$. Models assuming a standard
solar neighbourhood IMF were found to underpredict the numbers of
sub-mm galaxies by an order of magnitude, if these models were also
constrained to reproduce the present-day galaxy luminosity functions.

In the present study, we want to investigate whether a model that is
consistent with the new measurements of the Lyman-dropout galaxy
population at $z>6$ can produce sufficient ionizing photons to
reionize the universe at $z_{\rm reion} \gtrsim 10$. If so, we want to
quantify which galaxies dominate $\epsilon$, and which aspects of the
model affect $\epsilon$ most. A similar analysis based on the
\galform\ model was performed by \citet{Benson06} (see also
\citealt{Benson02a, Benson02b}), who also included a simple model for
the evolution of the H{\sc II} volume filling factor. The present
paper looks in more detail at the properties of the galaxies that
cause reionization and the dark matter halos that host them, and how
these are connected to the newly discovered $z>6$ drop-outs.  The
properties of Lyman-break galaxies predicted by \galform\ over the
whole redshift range $z=3-20$ have been analyzed in more detail in a
companion paper by \citet{Lacey10b}, which also makes a more detailed
comparison with the observed far-UV luminosity functions. We
emphasize that the default values of the \galform\ model parameters
used in the present work are identical to those chosen by
\citet{Baugh05}, which were adjusted to match a range of observed
galaxy properties at lower redshifts. We will couple the \galform\
source model with a radiative transfer scheme to investigate the
progression of reionization in more detail in a follow-up paper
\citep{Raicevic10}.

This paper is organised as follows. In Section 2 we briefly discuss
the main ingredients of \galform, paying particular attention to those
aspects that most affect the ionizing luminosities of the galaxies.
In Section~3 we show the evolution of the emissivity $\epsilon$ for
the default \galform\ parameters, discuss which galaxies dominate
$\epsilon$, and how changes in \galform\ parameters affect
$\epsilon$. In section 4 we show the corresponding far-UV luminosity
functions at $z=6$ and 10, explore the extent to which the currently
detected galaxies constrain $\epsilon$, and how future observations
with {\em e.g.} the {\em James Webb Telescope} will improve our
understanding.  We summarize in Section 5.

\section{Method}
\label{sect:model}
The \galform\ semi-analytical model \citep{Cole00} computes how
galaxies form and evolve in the hierarchically growing dark matter
halos of a cold dark matter Universe. The evolution of the halos
themselves is described by halo merger trees, which are either
extracted from an N-body simulation or computed using a Monte-Carlo
scheme based on \cite{lacey93} and improved by \cite{Parkinson08}. The
semi-analytical algorithm incorporates physically motivated recipes
for gas cooling, star formation, feedback from supernovae, galaxy
mergers, metal enrichment, dust production and other processes, and in
particular allows a calculation of the observable properties of each
galaxy, notably its broad-band luminosity and colours, and its
ionizing emissivity; see \cite{Baugh06} for a recent review of
semi-analytical methods.

The buildup of dark matter halos of course depends on the assumed
cosmological parameters, but the properties of the {\em galaxies}
associated with them are at least equally strongly dependent on the
\lq gastrophysics\rq\ governed by \galform\ parameters; for this
reason we only consider the cosmological parameters used in the
Millennium simulation \cite{Springel05-1}, $(\Omega_{\rm
m},\Omega_\Lambda,\Omega_{\rm b},h,\sigma_8,n_s) =
(0.25,0.75,0.045,0.73,0.9,1)$\footnote{Note that the original
\citet{Baugh05} model used a slightly different cosmology,
$(\Omega_{\rm m},\Omega_\Lambda,\Omega_{\rm b},h,\sigma_8,n_s) =
(0.3,0.7,0.04,0.7,0.9,1)$. The change of cosmological parameters was
introduced for consistency with the Millennium-II simulation
\citep{Boylan-Kolchin09} which we will employ in future numerical
simulations of reionization.}.

Even at redshift $z=0$ only a very small fraction of baryons have been
converted into stars \citep{Fukugita98}. In particular the faint-end
slope of the $z=0$ K-band luminosity function, $\alpha_{\rm L}\approx
-1$ \citep{Cole01-2}, is much flatter than the low-mass slope of the
dark halo mass function, $\alpha_{\rm M}\approx -2$
\citep{press74}. Therefore a crucial ingredient of any successful
galaxy formation model is strong negative feedback to quench the
formation of small galaxies \citep{White91,Benson03}. \galform\ incorporates
this and other effects with a set of rules, each with an associated
set of parameters. Some of these have a large effect on the properties
of early galaxies, others mostly affect the present-day galaxy
population. Recent studies using \galform\ have concentrated on two
different variants, that of \citet{Baugh05} (hereafter, \Baugh ) and
of \cite{Bower06} (hereafter, \Bower ), which adopt somewhat different
prescriptions for star formation, feedback and the IMF (see also
\citet{Lacey08} for more details about the \Baugh\ model). The \Baugh\
model includes superwinds (following \cite{Benson03}) in order to
better reproduce the bright-end of the optical and near-IR galaxy
luminosity function at $z=0$, while the \Bower\ model instead
accomplishes this by including feedback from accreting black holes
(see also \cite{Croton06}). The other most important difference
between the two models is that the \Baugh\ model assumes that stars
form with a top-heavy initial mass function (IMF) in starbursts, and a
normal solar neighbourhood IMF in galaxy discs, while the \Bower\ model
instead assumes that all star formation occurs with a solar
neighbourhood IMF. In addition to this, the two models make somewhat
different assumptions about the star formation timescale in discs,
supernova feedback, the timescale for ejected gas to be
re-incorporated into halos, and the triggering of starbursts. 

While the \Baugh\ and \Bower\ models predict similar galaxy luminosity
functions at optical and near-IR wavelengths at $z=0$, the \Baugh\
model is in much better agreement with the observed numbers of
star-forming galaxies seen at high redshifts, selected either as
Lyman-dropouts or from their sub-mm emission
\citep{Baugh05,Lacey10b}. As we will show later, the \Baugh\ model
also predicts higher ionizing emissivities at high redshifts than the
\Bower\ model, and a correspondingly higher redshift of reionization,
in better agreement with current observational constraints. For these
reasons, we concentrate in this paper on predictions from the \Baugh\
model.  Not surprisingly, neither superwinds nor AGN greatly affect
the predictions for galaxies at $z \gtrsim 6$, since the massive
galaxies that are affected by these processes are extremely rare at
such early times. Nevertheless, we do find quite significant
differences between these two popular \galform\ variants in what they
predict for ionizing emissivities at $z \gtrsim 6$, which are related
to their different assumptions about star formation, supernova
feedback and the IMF.  We now discuss the physical processes
incorporated in the \Baugh\ version of \galform\ model that have a
large effect on $z \gtrsim 6$ galaxies, and why they were introduced
in the original model.

\subsection{Star formation}
The model assumes two distict modes of star formation, {\em quiescent}
star formation in galaxy discs, and {\em starbursts} triggered by
galaxy mergers. In both cases the instantaneous star formation rate is
parametrized as:
\begin{equation}
\psi={M_{\rm cold}\over \tau_\star}\,,
\label{eq:sfr}
\end{equation}
where $M_\mr{cold}$ is the amount of cold gas in the galaxy, and
$\tau_\star$ the star formation time scale. Neglecting the life-times
of massive stars (the instantaneous recycling approximation), the
stellar mass in long-lived stars then builds up at a rate
\begin{equation}
  \dot{M}_\star = (1-R) \psi\,,
\end{equation}
where $R$ is the recycling fraction, see \cite{Cole00} for more details.  \\

\noindent In the {\em quiescent star formation mode}, $\tau_\star$
depends on the circular velocity, $V_\mr{disc}$, of the galactic disc
at the half-mass radius, as $\tau_\star=\tau_{\star,0}
(V_\mr{disc}/200~{\rm km\, s}^{-1})^{\alpha_\star}$, with
$\tau_{\star,0}=8$~Gyr and $\alpha_\star=-3$. This parametrization
yields reasonable gas masses and star formation rates at low redshifts
$z\sim 0$, and implies that $\psi$ is quite low at high
redshifts. This makes the high-$z$ discs gas rich, so that when
galaxies merge, there is a large reservoir of gas available for
fueling a starburst \citep{Baugh05}. \\

\noindent{\em Bursts of star formation} are assumed to be triggered by
galaxy mergers under certain conditions. The model includes both {\em
  major} and {\em minor} mergers, distinguished by the mass ratio of
merging galaxies.  Major mergers between spirals are assumed to
destroy both discs and consume the remaining gas in a starburst. Minor mergers were introduced in the model motivated by the simulations of \cite{Hernquist95}; such a merger does not destroy the disc, but does build up the bulge. The star formation time scale in the burst mode is shorter than in the quiescent mode
\citep[see][]{Baugh05}.\\

\noindent{\em The stellar initial mass function} for quiescently
forming stars is assumed to be similar to what is observed in the
solar neighbourhood, specifically that proposed by \cite{Kennicutt83},
$dN/d\ln(m) \propto m^{-x}$, with $x=0.4$ for $m <1M_\odot$ and
$x=1.5$ for $m > 1M_\odot$, However, in bursts the IMF is assumed to be top-heavy, $x = 0$. In both cases, the IMF covers the mass range $0.15 < m/M_\odot < 120$. \\

Star formation with a top-heavy IMF in bursts triggered by gas-rich
galaxy mergers results in large UV luminosities from the massive young
stars, and also the production of large quantities of metals and dust
from supernovae. This dust in turn absorbs the copious UV radiation
and re-radiates it at far-IR wavelengths. Both the frequent bursts at
high redshifts and the top-heavy IMF are needed to boost the number of
very luminous high-z IR galaxies to a level consistent with the
observed number counts and redshift distribution of sub-mm
galaxies. The parameters in the \Baugh\ model were chosen to match
this sub-mm data, while at the same time yielding good fits to the
Lyman-break galaxy luminosity function at $z\sim 3$, and remaining
conistent with observational constraints at $z=0$ \citep{Baugh05}. The
case for a top-heavy IMF for the formation of at least a fraction of
stars is further supported by the fact that its use during starbursts
also results in better agreement with observed metallicities
(including $\alpha$/Fe ratios) in intracluster gas in clusters and
stars in elliptical galaxies \citep{Nagashima05a,
Nagashima05b}. Other independent observational evidence for
variations in the IMF is discussed in \citet{Lacey10a,Lacey10b}. We
emphasize that our results do not depend crucially on the precise form
of the top-heavy IMF assumed - similar results would be obtained for
an IMF in which the high-mass slope was fixed but the low mass
turnover was varied, as proposed by \citet{Larson98}.  We will show
below that the bursts, and the associated change in the IMF during
bursts, {\em both} have large effects on the emissivity of ionizing
photons by \galform\ galaxies at $z \gtrsim 6$.

\subsection{Supernova feedback}
\label{sect:SNe}
The fact that galaxies in low-mass halos form stars very inefficiently
is likely due to energy injection from supernovae \citep{Dekel86}. In
the \Baugh\ model this is implemented by ejecting gas out of a
galaxy disc at a rate 
\begin{equation}
\dot M_{\rm eject}=\psi \left({V_{\rm disc}\over V_{\rm
      hot}}\right)^{-\alpha_{\rm hot}}\,,
\label{eq:SNfeedback}
\end{equation}
so that it is no longer available for star formation. Here, $V_{\rm
disc}$ is the circular velocity of the galactic disc at the half-mass
radius. Values of $V_{\rm hot}=300$~km~s$^{-1}$ and $\alpha_{\rm
hot}=2$ were chosen to reproduce the faint-end slope of the $B$-band
galaxy luminosity function at $z=0$ \citep{Baugh05}. Such strong
feedback also significantly quenches star formation in small halos at
$z \gtrsim 6$, and therefore has a large impact on reionization. Note
that the \Bower\ model incorporates even stronger SN feedback in small
halos.

\subsection{Photo-ionization feedback}
\label{sect:photo}
Star formation in small galaxies may be quenched as the IGM becomes
ionized, either because cooling is suppressed \citep{Efstathiou92}, or
because the higher IGM gas pressure inhibits gas from falling into halos
\citep{Gnedin00b}, or because photo-heating causes small galaxies to
lose their gas \citep{Hoeft06,Okamoto08}. These effects may lead to a
global suppression of star formation during and after the EOR, as seen
in the simulations of \cite{Crain09}.  The standard approach in
\galform\ is to model this by suppressing the cooling of halo gas onto
the galaxy when the host halo circular velocity is below a threshold value
\begin{equation}
V_{\rm circ} < V_{\rm cut}\,.
\label{eq:photo}
\end{equation}
at redshifts $z<z_{\rm cut}$ (but see also \citet{Benson02a} for a
more detailed treatment). 

The default value of $V_{\rm cut}=60$~km~s$^{-1}$ in the \Baugh\
model, originally guided by the results of \cite{Gnedin00b}, is
considerably larger than values found from more recent simulations
\citep{Hoeft06,Okamoto08}. The original \Baugh\ model also assumed
$z_{\rm cut}=6$. Interestingly, because only the {\em gas cooling} in
the halo is suppressed in \galform, a small galaxy with circular
velocity $V<V_{\rm cut}$ can continue to form stars until it has
exhausted its supply of {\em cold} ({\em i.e.}  already cooled)
gas. This way of suppressing galaxy formation in small halos once the
IGM is ionized has consequences for reionization and also for the
luminosity function at later times, as we show below.

\subsection{Modeling Lyman-continuum and broad-band SEDs}
\label{sect:SED}
The \galform\ code computes the spectral energy distribution (SED) of
each galaxy, given its star formation history and abundance
evolution. The population synthesis models are based on the Padova
stellar evolution tracks combined with Kurucz model atmospheres
\citep{Bressan98}. The dust extinction is modeled with a prescription
described by \cite{Cole00} with improvements described in
\cite{Lacey10b}. Convolving the SED with a filter response yields
broad-band luminosities for the galaxy.
Below we will use the rest-frame 1500\AA\ broad-band AB magnitudes of
\galform\ galaxies to compare against observed galaxy luminosity
functions at approximately the same rest-frame wavelength, after
rescaling observed luminosities and number densities to the same
\mill\ cosmology as assumed in the model.

\galform\ also computes the Lyman-continuum luminosity for each
galaxy, expressed as the emission rate of ionizing photons,
\begin{equation}
\dot N_{\rm LyC} = \int_{\nu_{\rm thresh}}^\infty {L_{\nu} \over h\nu}\,d\nu\,,
\end{equation}
where $L_{\nu}$ is the SED of the galaxy and $\nu_\mr{thresh}$ is the
Lyman-limit frequency, $h \nu_\mr{thresh} = 13.6$ eV.  Note that the
number of ionizing photons produced per solar mass of stars formed is
very different for the Kennicutt IMF assumed during quiescent star
formation compared to the top-heavy IMF in bursts ($N_{\rm
LyC}/M_\star=3.2 \times 10^{60}$ and $3.5\times 10^{61}$ respectively,
for solar metallicity).

A considerable fraction of those ionizing photons may be absorbed
locally in the interstellar medium of the galaxy or by gas in the
surrounding halo, and the fraction $f_{\rm esc}$ of photons that does
manage to escape into the IGM is very uncertain. Observations of
$z\sim 3-4$ Lyman-break galaxies (LBGs) by \cite{Steidel01} and
\cite{Shapley06} suggest $f_\mr{esc}\sim 0.01-0.1$ or even lower
\citep{Giallongo02} (but note the slightly different definition of
$f_{\rm esc}$ there).  The escape fraction may depend strongly on the
porosity of the interstellar medium within the galaxy or the presence
of supernova-driven winds \citep[e.g.][]{ciardi02, clarke02}. Some of
the more recent models that attempt to include these effects suggest
that $z \gtrsim 6$ galaxies may have significantly larger escape
fractions, $f_\mr{esc} \sim 0.5$, \citep[e.g.][]{Wise09, Razoumov10}.
Note that this parameter is unlikely to be independent of metallicity,
gas content, and halo mass. In this paper, we simply assume $f_\mr{esc}$ to be the same for all galaxies.

We will now discuss the net emissivity of ionizing photons in the
\Baugh\ model, and how that depends on \galform\ parameters.

\section{Ionizing emissivities}

\begin{figure}
  \begin{center}
    \includegraphics[width=0.45\textwidth,keepaspectratio = true, clip=true, trim=0 30 20 30]{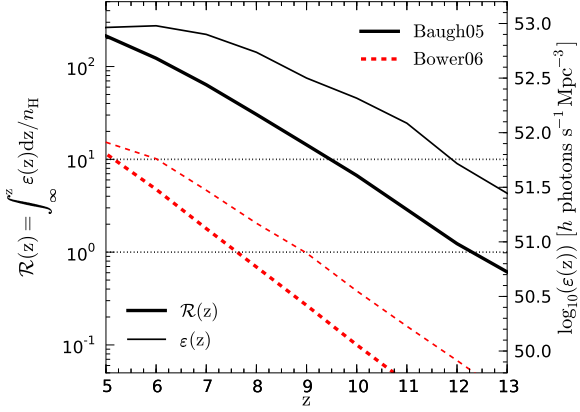}
  \end{center}
  \caption{The ratio ${\cal R}(z)$ of the number of ionizing photons
    produced per hydrogen atom up to redshift $z$ in the two fiducial
    \galform\ models, \Baugh\ and \Bower\ (thick lines, left y-axis)
    as well as the total emissivity, $\epsilon (z)$, in the same
    models (thin lines, right y-axis). The horizontal dashed lines
    mark the minimum number of photons per H atom that must be
    produced to achieve reionization: in the most optimistic case,
    only one (bottom line), but 10 or more when reasonable values for
    the ionizing escape fraction and mean number of recombinations per
    H atom are taken into account (top line). The \Baugh\ model
    produces $\sim$ 100 times more ionizing photons at $z \sim
    10$ than \Bower\ and reaches 10 photons per H atom $\Delta z
    \sim 5$ earlier. The decreased slope in $\epsilon(z)$ at $z
    \leq 6$ is caused by the turn-on of photo-ionization feedback at
    $z=6$ in both models. }
  \label{fig:bau_bow_nphot}
\end{figure}

The emissivity $\epsilon(z)$, the number of ionizing photons produced
per unit comoving volume at redshift $z$, is found by summing the
Lyman-continuum luminosity of all galaxies, per unit volume,
\begin{equation}
\epsilon(z) = \int_0^{\infty}\,\dot N_{\rm LyC}\,\Phi(\dot{N}_{\rm LyC})\,\dd \dot{N}_{\rm LyC},
\end{equation}
where $\Phi(\dot{N}_{\rm LyC})$ is the Lyman-continuum luminosity
function. The emissivity $\epsilon(z)$ increases by approximately
1.5~dex between $z=13$ and $z=5$ in the \Baugh\ model
(Fig.\ref{fig:bau_bow_nphot}, thin line), mostly as a consequence of
evolution in the halo mass function, as we will show below.

Integrating $\epsilon(z)$ down to a given redshift yields the total
number of ionizing photons produced per unit comoving volume up to that
time. This number can be compared to the mean comoving number density
of hydrogen atoms, $n_{\rm H}$.  Reionization will
occur when their ratio
\begin{equation}
{\cal R}(z)\equiv {\int_\infty^z\,\epsilon(z)\,dz\over n_{\rm H}\,},
\end{equation}
is ${\cal R}=(1+N_{\rm rec})/f_{\rm esc}$. Here, $N_{\rm rec}$ denotes
the mean number of recombinations per hydrogen atom up to
reionization, and $f_{\rm esc}$ is the mean escape fraction from
Section \ref{sect:SED}.

Estimating $N_{\rm rec}$ is not straightforward. Recombinations can
occur in the higher-density regions of the general IGM, in \lq
mini-halos\rq\ that have too shallow potential wells for star
formation \citep{Shapiro04,Ciardi06}, or in even higher-density
regions associated with Lyman-limit or damped Lyman-$\alpha$ systems.
The value of $N_{\rm rec}$ will itself depend on
$\int_\infty^z\,\epsilon(z)\,dz$, since a slower build-up of the
ionization rate will allow more time for
recombinations. Interestingly, once the IGM is ionized, the smoothing
of the density field due to gas pressure following photo-heating
reduces the recombination rate \citep{Pawlik09}. Current simulations
of the EoR suggest values of $N_{\rm rec}$ of a few \citep{Iliev06,
McQuinn07, Trac07}.

Combining the estimate of $1+N_{\rm rec}\sim 2$ with a reasonable
escape fraction of $f_{\rm esc}\sim 0.2$ then suggests that
reionization requires a value of ${\cal R}\sim 10$. This is plotted as
a function of redshift for the default values of the \Baugh\ and
\Bower\ \galform\ parameters in Fig.~\ref{fig:bau_bow_nphot} (thick
lines), suggesting that the \Baugh\ model will produce a reasonable
reionization redshift $z_{\rm reion}\sim 10$, $\Delta z\sim 5$ before
\Bower . 
Next we discuss the properties of the galaxies and halos
that dominate the emissivity in the \Baugh\ model, and how strongly
these depend on the assumed parametrization in the model, following
the same order as in the previous Section~\ref{sect:model}.

\subsection{Effect of star formation parameters and IMF}

\begin{figure}
  \begin{center}
    \includegraphics[width=0.45\textwidth,keepaspectratio = true,clip=true, trim = 0 10 20 30]{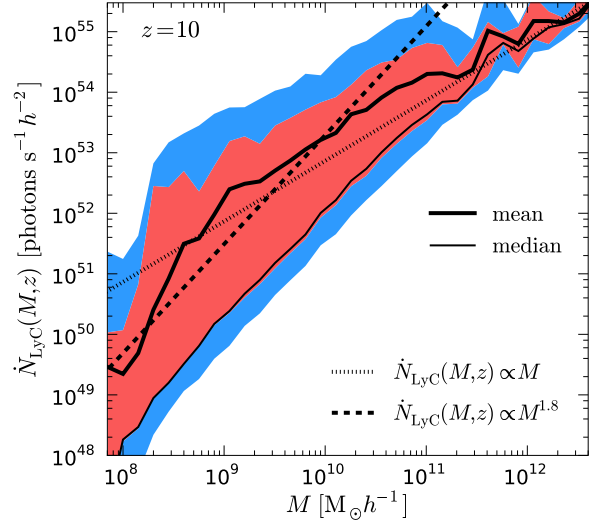}
  \end{center}
  \caption{Lyman-continuum photon luminosity, $\dot N_{\rm {LyC}}(M,z)$,
    of halos as a function of halo mass $M$, in the \Baugh\ model at
    $z=10$ (median and mean relation are shown as thick and thin solid
    lines, respectively).  $\dot N_{\rm {LyC}}$ increases approximately
    as $\dot{N}_{\rm{LyC}} \propto M^{1.8}$ for small halos $M\lesssim
    2\times 10^9\,h^{-1}M_\odot$, and as $\dot{N}_{\rm{LyC}} \propto M$
    for more massive halos, with little dependence on redshift. 
   The 50 and 90 per cent ranges of $\dot N_{\rm LyC}(M)$ at given halo
    mass are shaded red and purple, respectively.  There is up to
    5~dex range in $\dot N_{\rm LyC}$ at a given mass, a consequence
    of the dominance of starbursts in producing ionizing photons.}
\label{fig:emis_evol2}
\end{figure}

\begin{figure}
  \begin{center}
    \includegraphics[width=0.45\textwidth,keepaspectratio = true,clip=true, trim=0 10 20 25]{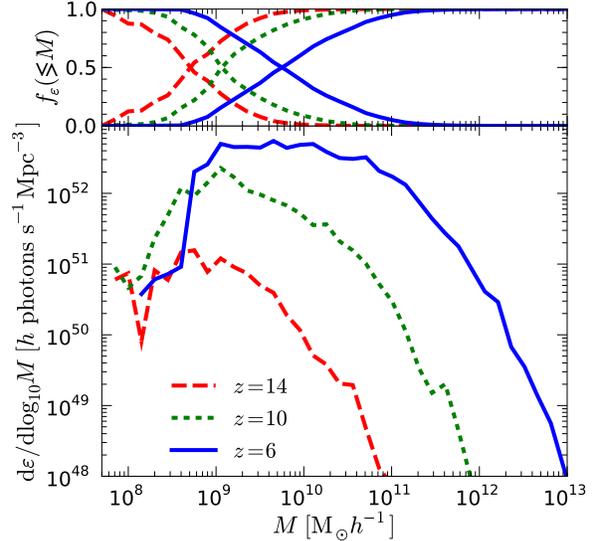}
  \end{center}
  \caption{{\em Main panel}: Lyman-continuum emissivity as a function
    of halo mass, $\dd \epsilon(M,z)/ \dd \log_{10}(M)$, for various
    redshifts indicated in the panel. The emissivity is low for very
    low-mass halos that are unable to cool gas, reaches a peak which
    increases with decreasing $z$, and a tail towards larger masses
    set by the exponential drop in the number of massive halos. At
    $z\sim 10$ most ionizing photons are produced by halos in a
    relative small mass range, $\sim 1$~dex.  {\em Top inset}:
    cumulative fraction $f_c$ of ionizing photons produced in halos
    more massive or less massive than a given value (rising and
    falling curves, respectively). The mass of halos below which 50
    per cent of ionizing photons is produced rises by approximately an
    order of magnitude from $\sim 8\times 10^8\,\msun$ at $z=14$ to
    $\sim 8\times 10^9\, \msun$ at $z=6$.}
  \label{fig:emis_evol3}
\end{figure}

\begin{figure}
  \begin{center}
    \includegraphics[width=0.45\textwidth,keepaspectratio = true, clip=true, trim = 0 20 20 30]{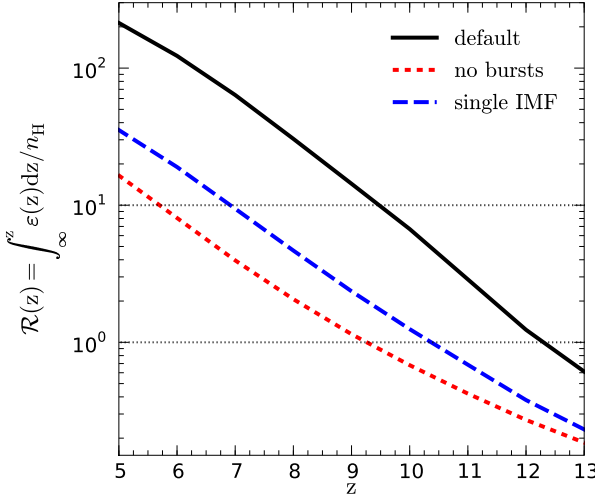}
  \end{center}
  \caption{Dependence of the total number of ionizing photons produced
    per hydrogen atom up to redshift $z$, ${\cal R}(z)$, on the
    starburst parameters in {\Baugh}: default model (black), no bursts (red), including bursts, but not the change
    to a top-heavy IMF in bursts (blue). Including bursts increases
    $\epsilon(z)$ by a factor 5-10, depending on redshift. The effect
    of the change in IMF in the bursts is large, yet even without it
    bursts still increase $\epsilon$ by a factor of $\sim
    2$. Neglecting bursts delays reionization (${\cal R} = 10$) by
    $\Delta z \sim 4$.}
\label{fig:bursts_nphot}
\end{figure}

\begin{figure}
  \begin{center}
    \includegraphics[width=0.45\textwidth,keepaspectratio = true, clip=true, trim=0 10 20 25]{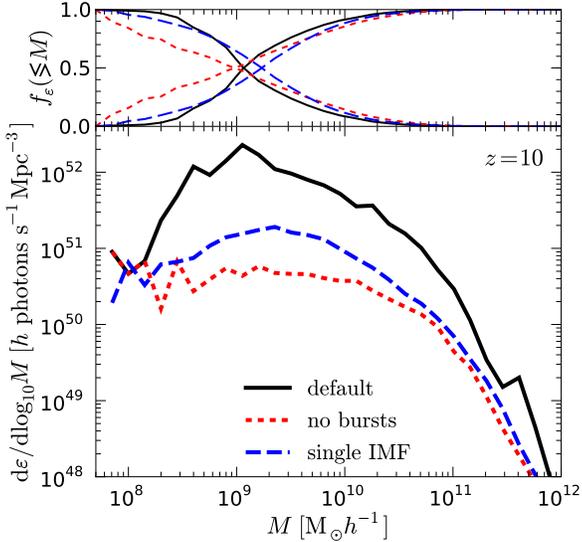}
  \end{center}
  \caption{Dependence of emissivity as a function of halo mass, $\dd
    \epsilon / \dd \log_{10} M$, on the burst parameters in the
    \Baugh\ model. The characteristic halo mass at which 50 per cent
    of the ionizing photons is produced does not greatly depend on the
    burst parameters. However, switching off the bursts (red short
    dashed line) extends the halo mass range in which the majority ($
    \sim$ 90 per cent) of ionizing photons is produced by $\sim$
    1 order magnitude in comparison to the default model (solid black
    line).}
  \label{fig:bursts}
\end{figure}

\begin{figure}
  \begin{center}
    \includegraphics[width=0.45\textwidth,keepaspectratio = true, clip=true, trim=20 10 40 25]{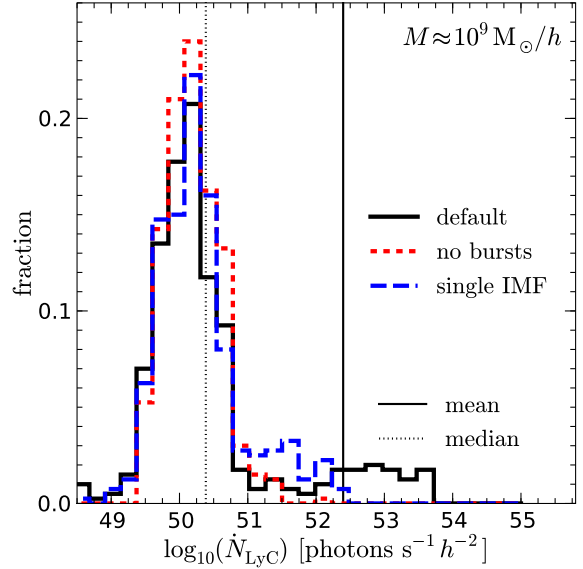}
  \end{center}
  \caption{Distribution of Lyman-continuum photon luminosities, $\dot
    N_{\rm {LyC}}$ at $z=10$, for halos with mass $M \approx 10^9
    \msun$. Different line styles refer to different models for the
    bursts, vertical dotted and solid lines indicate median and mean
    $\dot N_{\rm {LyC}}$ in the default model, respectively.  The distribution of $\dot
    N_{\rm {LyC}}$ peaks at a few times $10^{50}$~$h^{-2}$ photons
    ~s$^{-1}$, but allowing bursts introduces a long tail towards much
    more luminous galaxies (red versus black histograms), with the
    change in IMF in bursts having a large contribution to this (blue
    versus black histograms). This tail makes the mean $\dot N_{\rm
    {LyC}}$ nearly 2~dex brighter than the median.  In the default
    model with a top-heavy IMF in bursts there is a nearly 5~dex range
    in luminosity at given halo mass.}
\label{fig:bursts_hist}
\end{figure}

The number of ionizing photons produced per unit time by galaxies in a
halo of given mass, $\dot N_{\rm LyC}(M,z)$, is plotted as a function
of $M$ in Fig.~\ref{fig:emis_evol2}. The virial temperature $T_{\rm
vir}$ of halos with $M<M_{\rm min} \approx 10^8\,h^{-1}M_\odot$ is too
low to enable radiative cooling by atomic lines and hence such halos
do not form stars\footnote{We recall that this \galform\ model does
not consider Pop.~III stars that form due to molecular cooling in such
small halos.}.  Given that $T_{\rm vir} \propto (1+z)$ at fixed $M$,
there is strong redshift dependence in $\dot{N}_{\rm LyC}(M,z)$ at
very low masses, but above this minimum mass \galform\ predicts
essentially no evolution in the mean $\dot{N}_{\rm LyC}(M,z)$ between
$z=15$ and $z=6$, but with a modest $\sim50\%$ decrease in the median
in halos with mass $M \gtrsim 10^{10} \msun$ in the same redshift
range.

The mean $\dot N_{\rm {LyC}}$ at a given halo mass increases
approximately as $\dot{N}_{\rm{LyC}} \propto M^{1.8}$ for small halos
$M\lesssim 2\times 10^9\,h^{-1}M_\odot$, and roughly as
$\dot{N}_\mr{LyC} \propto M$ for more massive halos, in contrast to
many recent simulations of reionization which assume a simple $\dot
N_{\rm {LyC}}\propto M$ relation for all $M$
\citep[e.g.][]{Furlanetto04, Iliev06}.  Interestingly, there is a very
large difference between the mean and median of $\dot{N}_{\rm LyC}$ at
given $M$, and there is also a very large range, up to $\sim 5~$dex, in
$\dot N_{\rm {LyC}}$ at {\em given} $M$ (Fig.~\ref{fig:emis_evol2}).
Both are consequences of the importance of bursts in generating
ionizing photons, as we will discuss in more detail below.

The total Lyman-continuum emissivity per dex in halo mass $\dd\epsilon/ \dd\log_{10}(M)$ 
(Fig.~\ref{fig:emis_evol3}), can be obtained by combining the mean luminosity of a single halo of given mass, $\dot N_{\rm
{LyC}}(M)$, with the number of halos of that mass, $\dd
n/\dd\log_{10}(M)$. This function evolves rapidly as a
consequence of the rapid build-up of more massive halos as time
progresses.  The halo mass below which 50 per cent of ionizing
photons are produced increases from $\sim 8\times 10^8\,h^{-1}M_\odot$
at $z=14$ by an order of magnitude to $\sim 8\times
10^9\,h^{-1}\,M_\odot$ at $z=6$ (top panel of
Fig.~\ref{fig:emis_evol3}). At high $z$, the mass range of halos that
contribute significantly to $\epsilon$ is relatively small, of order
1~dex, since it is limited at low $M$ by $M_{\rm min}$ and at large
$M$ by the exponential drop in the abundance of more massive halos.
At later redshift $z\sim 6$, $\dd \epsilon/ \dd {\rm log_{10}} M$ is
nearly independent of $M$ over nearly 2~dex, a consequence of the fact
that the ionizing photon luminosity of halos increases with halo mass
approximately as $\dot N_{\rm {LyC}}(M)\propto M^1$ (dotted line in
Fig.~\ref{fig:emis_evol2}), whereas the number density of halos
decreases with increasing mass approximately as $\dd n/ \dd
\log_{10}M\propto M^{-1}$.

The impact of starbursts on the emissivity is quantified in
Fig.~\ref{fig:bursts_nphot}. In the default \Baugh\ model, bursts
increase the ionizing emissivity relative to that from quiescent
galaxies both as a consequence of the reduction in star formation
timescale, Eq. \refeq{eq:sfr}, and because of the assumed change to a
top-heavy IMF. The net effect is a factor 5-10 increase in $\epsilon$
depending on redshift, with approximately 65 per cent of the increase
due to bursts following a minor merger.  Most of the increase in $\dot
N_{\rm {LyC}}$ is a consequence of the assumed change in IMF.

Neglecting bursts does not affect the \lq characteristic\rq\ halo mass
below which 50 per cent of the ionizing photons are produced
(Fig.~\ref{fig:bursts}) but it does increase the range of halo masses
responsible for the majority (e.g. 90 per cent) of ionizing photon
production by $\sim$ 1 dex (compare solid black and short dashed
red lines in the top inset of the same panel). 

Bursts skew the distribution of $\dot N_{\rm {LyC}}$ at given halo
mass by introducing a long tail of much more luminous galaxies which
happen to be bursting, with again the assumed change in IMF playing a
dominant role (Fig.~\ref{fig:bursts_hist}). These few, but relatively
bright, galaxies dominate the emissivity at that halo mass by a large
factor. Remarkably, there can be nearly a 5~dex range in
Lyman-continuum luminosity at given halo mass.

We conclude that bursts are a crucial ingredient in order for the
\Baugh\ model to produce that many ionizing photons by $z\sim 10$. Not
only do stars form at a greater rate due to the decrease in the star
formation timescale, but especially the change to a top-heavy IMF in
bursts, originally introduced to produce sufficiently luminous sub-mm
galaxies at $z=1$-3, and to produce sufficient metals by $z=0$, causes
a small fraction of galaxies to emit copious ionizing radiation.  The
bursts occur mostly due to minor mergers, and are so effective because
the merging galaxies are very gas rich, itself a consequence of the
inefficient star formation in their quiescent state. Bursts also
introduce nearly 5~dex of scatter in the $\dot N_{\rm {LyC}}$-halo
mass relation.  These same bursts are also a crucial ingredient for
reproducing the observed luminosity function of Lyman-break galaxies
at $z>6$, as shown in \citet{Lacey10b} and also discussed below
(Fig.~\ref{fig:LF_bursts}).  But first we investigate the effect of
the feedback parameters on $\epsilon$.

\subsection{Effect of supernova feedback parameters}

\begin{figure}
  \begin{center}
    \includegraphics[width=0.45\textwidth,keepaspectratio = true, clip=true, trim=5 20 5 40]{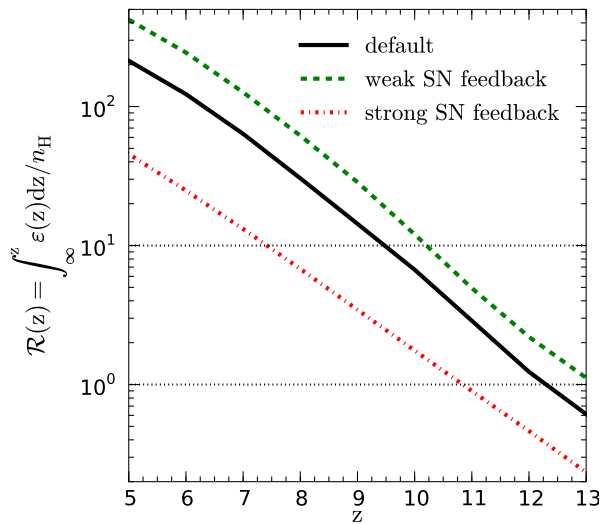}
  \end{center}
  \caption{Dependence of the total number of ionizing photons produced
    per hydrogen atom up to redshift $z$, ${\cal R}(z)$, on the
    \galform\ parameters that govern supernova feedback. Decreasing
    the efficiency of SN feedback (green dashed line) doubles the
    production of ionizing photons, resulting in reionization
    occurring $\Delta z \sim 0.7$ earlier. Increasing feedback from
    supernovae to the values used in the \Bower\ model (red dot-dashed
    line) delays reionization by $\Delta z \sim 2$.}
  \label{fig:SN_nphot}
\end{figure}

We consider two variants to the default \Baugh\ supernova feedback
parametrization to investigate how strongly they affect the emissivity
of ionizing photons. The \lq weak\rq\ feedback choice, shown in
Fig.~\ref{fig:SN_nphot} (green dashed line), uses parameters $(V_{\rm
hot},\alpha_{\rm hot})=(100\,{\rm km}~{\rm s}^{-1},1)$ (as defined in
Eq.~\ref{eq:SNfeedback}), as opposed to the default \Baugh\ values of
$(300\,{\rm km}~{\rm s}^{-1},2)$.  The ionizing emissivity of the weak
feedback model is not very different from a model without any SN
feedback at all; it produces nearly twice as many ionizing photons as
the default \Baugh\ model, increasing the reionization redshift, for
which ${\cal R}=10$, by $\Delta z\sim 0.7$. The \lq strong\rq\
feedback model has $(V_{\rm hot},\alpha_{\rm hot})=(500\,{\rm km}~{\rm
s}^{-1},3)$, close to the values $(475\,{\rm km}~{\rm s}^{-1},3.2)$
used in \Bower\ ; this choice of parameters decreases $\epsilon(z)$ by
a factor $\sim 5$, delaying reionization by $\Delta z \sim 2$.

Even stronger feedback is probably ruled out by the comparison with
the observed $z=6$ Lyman-break far-UV LF discussed in
Fig.~\ref{fig:SN_LF} below, but all three models are probably equally
consistent with the $z=10$ LF. This is not surprising since the SN
parameters affect mostly the fainter galaxies that are currently below
the detection limits at these very high redshifts. We note that the
standard approach in \galform\ modelling is to constrain the SN
feedback parameters by comparison with galaxy properties at
$z=0$. However, even if one chooses to relax the $z=0$ constraints on
the SN feedback, on the grounds that SN feedback might operate
differently in early galaxies, the constraints on this from the $z
\geq 6$ Lyman-break LFs still limit the uncertainty in $\epsilon$ to a
factor $\sim 2$ in the \Baugh\ model.

\subsection{Effect of photo-ionization feedback parameters}
\label{sect:photo}

\begin{figure*}
  \begin{center}
    \includegraphics[width=0.45\textwidth,keepaspectratio = true, clip=true, trim=0 0 20 0]{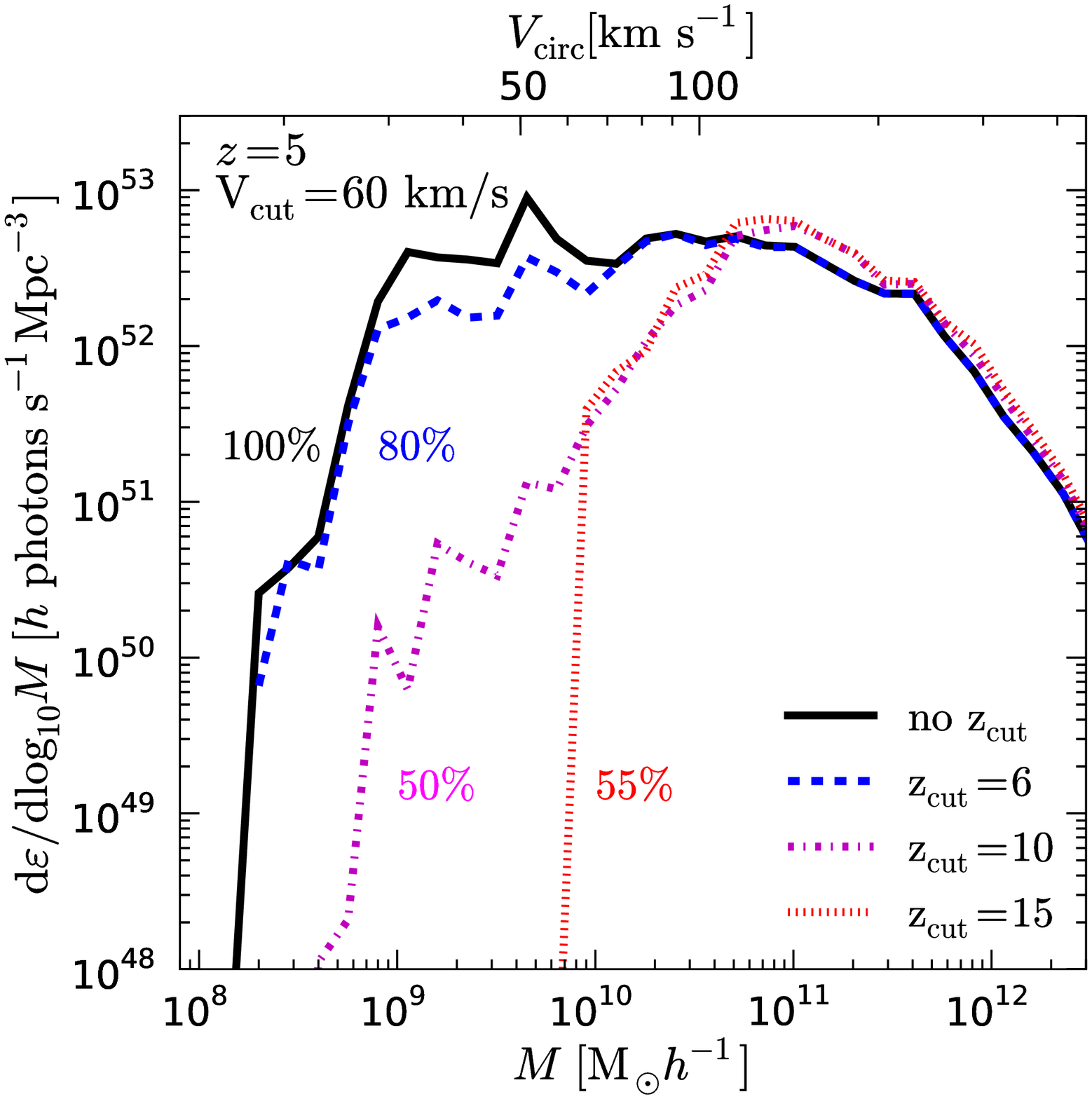}
    \includegraphics[width=0.45\textwidth,keepaspectratio = true, clip=true, trim=0 0 20 0]{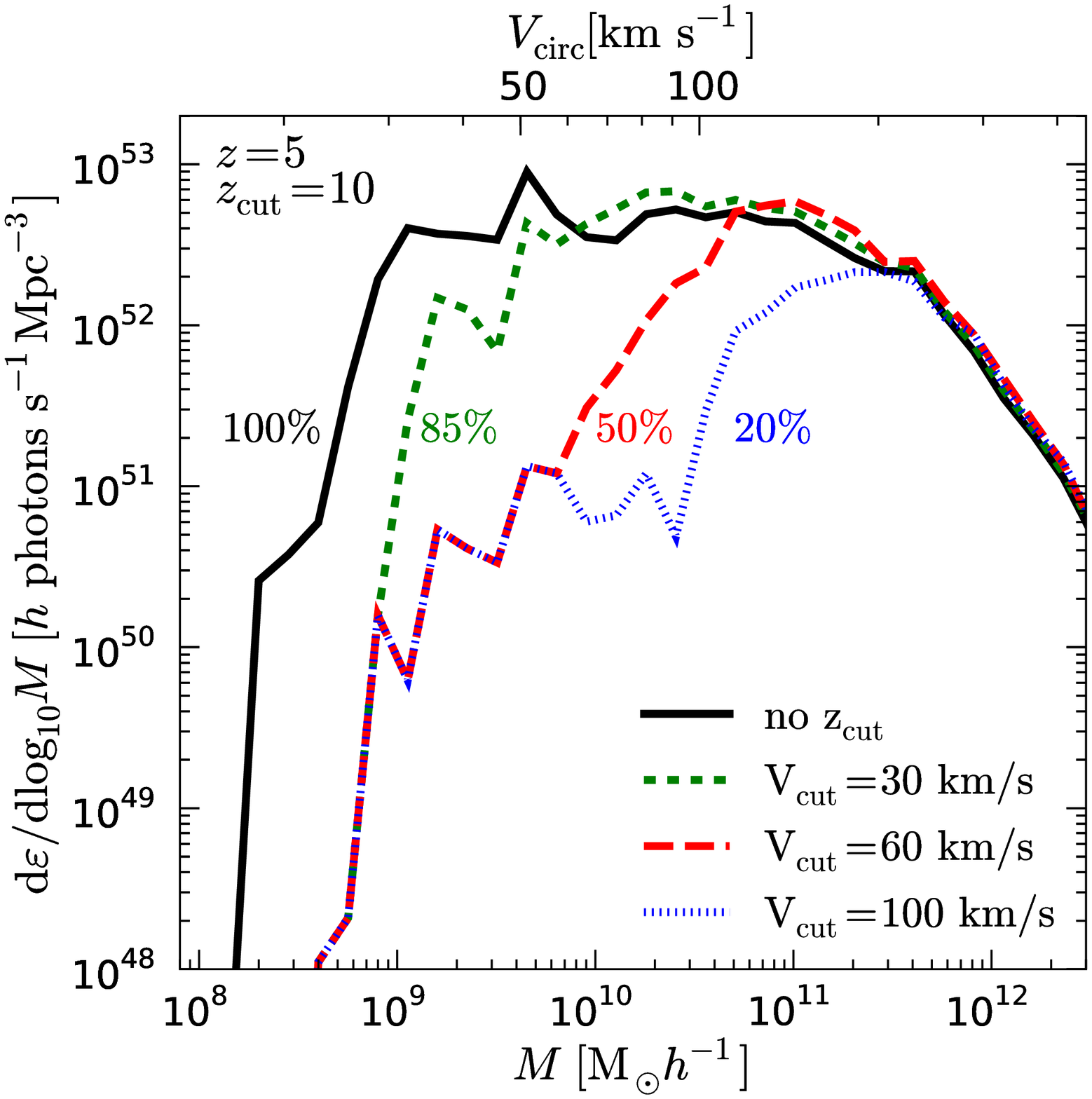}
  \end{center}
  \caption{Dependence of Lyman-continuum emissivity as a function of
    halo mass, $ \dd \epsilon/ \dd \log_{10}M$, at redshift $z=5$, on
    the \galform\ parameters that describe photo-ionization
    feedback. \textit{Left panel}: dependence on the reionization
    redshift $z_{\rm cut}$ (indicated in the legend; \lq no $z_{\rm
    cut}$\rq\ assumes reionization occurs below $z=5$) below which the
    IGM is assumed to be fully ionized. The impact of photo-ionization
    suppression takes a long time to take effect, but when suppression
    sets in it dramatically reduces the LC luminosity of the
    galaxies. \textit{Right panel}: dependence on the circular
    velocity $V_{\rm cut}$ below which photoionization feedback
    affects the galaxy. The suppression in $\dd \epsilon/ \dd
    \log_{10}M$ becomes evident at circular velocities below $\sim
    2V_{\rm cut}$. The scale on top of both panels gives the circular
    velocity of the halos (in $\rm km \, s^{-1}$) at $z=5$.  The
    numbers next to the lines give the ratio of the total emissivity
    of that model compared to the model with no photoionization
    feedback (black lines).}
  \label{fig:reion_params}
\end{figure*}

As discussed in Section \ref{sect:photo}, the effect of
photo-ionization feedback from reionization on galaxy formation is
modeled in \galform\ with a simple prescription, whereby gas cooling
is suppressed in all halos of circular velocity $V_\mr{circ} <
V_\mr{cut}$ after the reionization redshift $z_\mr{cut}$,
Eq.~(\ref{eq:photo}). The key feature of this prescription is that the
cold gas already present in galaxies before the onset of
photo-ionization feedback is allowed to form stars after
$z_\mr{cut}$. This results in a significant delay between the time at
which the surroundings of the galaxy become ionized and the quenching
of star formation. This is in contrast to several current simulations of
reionization, which assume that suppression is instantaneous
\citep[e.g.][]{Iliev06}. The delay is in fact so large that
the suppression of star formation (and hence also the production of
ionizing photons) due to photo-ionization has little effect on the
progression of reionization, as we will show elsewhere.

However, given enough time, photo-ionizing feedback does have a strong
effect on the ionizing emissivity, as shown in
Fig.~\ref{fig:reion_params}. Note that the default \Baugh\ model uses
a value of $V_{\rm cut}=60$~km~s$^{-1}$ which is unrealistically high
compared to more recent simulation results, which reduces $\epsilon$
by as much as 50 per cent by redshift 5 compared to the no
reionization model (assuming reionization occurs at $z_\mr{cut} =
10$).  The more modern value of $V_{\rm cut}\sim 30$~km~s$^{-1}$,
suggested by the simulations of \cite{Okamoto08}, yields a smaller yet
still significant decrease in the total emissivity at $z=5$ of 15 per
cent.

We conclude that photo-ionization suppression as implemented in
\galform\ has little effect on the production of ionizing photons
until well after reionization, but it does affect the emissivity at
later times. Interestingly, the photo-ionization quenching of star
formation also has observable effects on the Lyman-break LF, as we
discuss in more detail below (Fig.~\ref{fig:zcut_LF}).

\section{Far-UV luminosity functions of the galaxies that caused reionization}
\begin{figure}
  \begin{center}
    \includegraphics[width=0.45\textwidth,keepaspectratio=true, clip=true, trim=10 10 30 10]{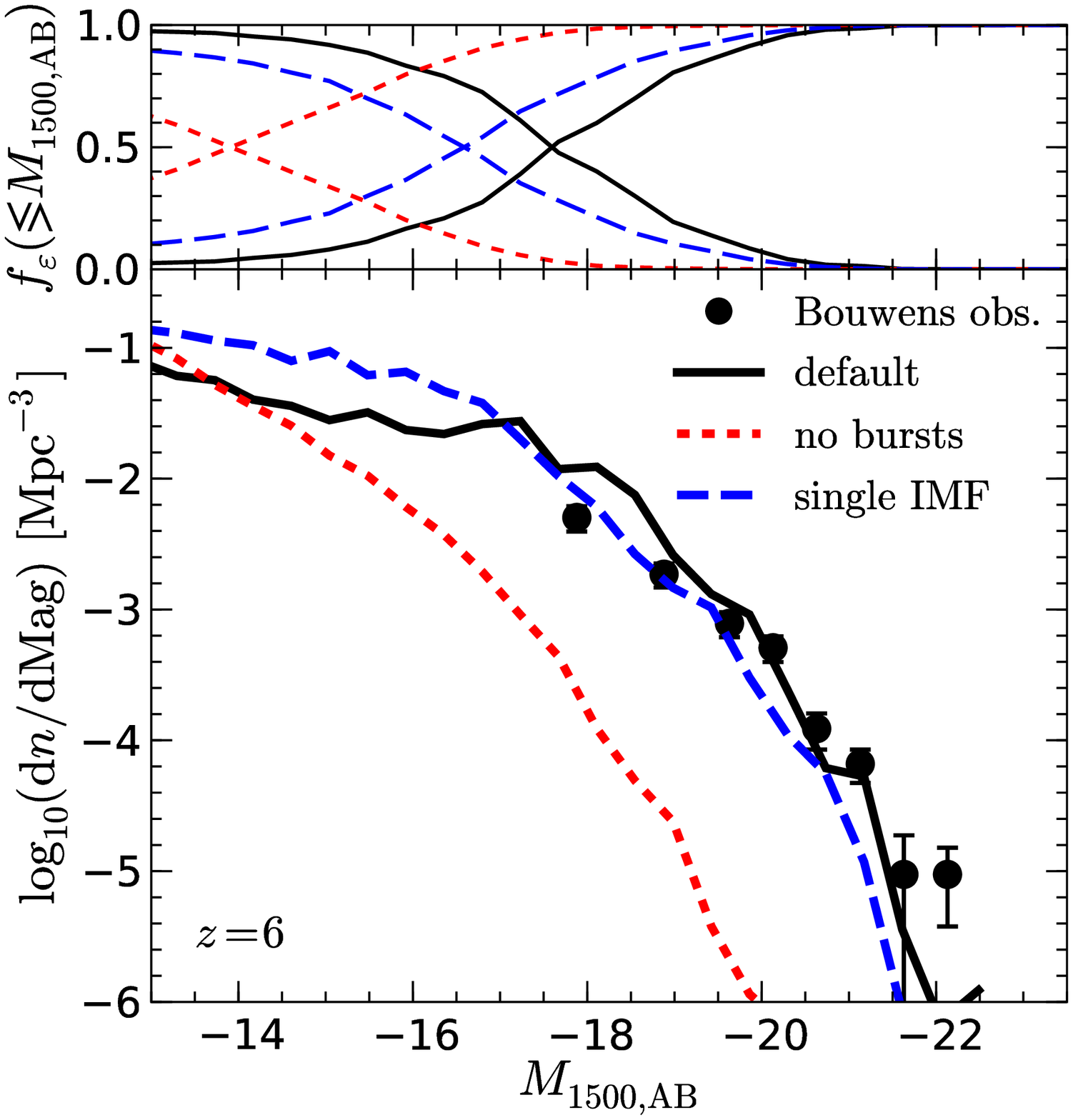}
    \includegraphics[width=0.45\textwidth,keepaspectratio=true, clip=true, trim=10 10 30 10]{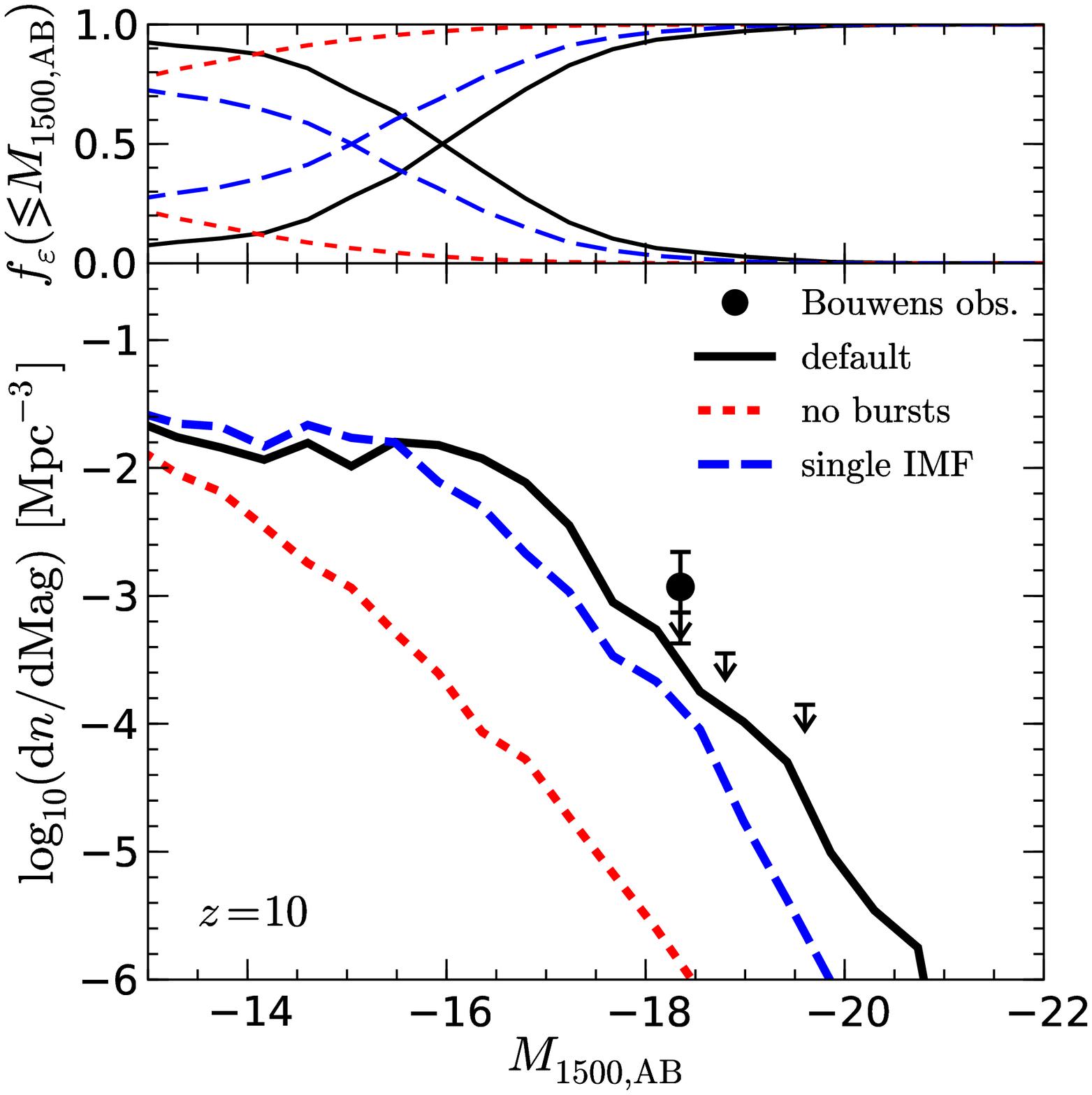}
  \end{center}
  \caption{Rest-frame 1500 \AA\ broad-band luminosity functions of the
    default \Baugh\ model (lines) compared to data from
    \protect\cite{Bouwens07} and \protect\cite{Bouwens09a}, at
    redshifts $z=6$ and 10 (symbols with error bars; downward pointing
    arrows mark $1 \sigma$ upper limits).  Both the default \Baugh\
    model (black solid lines) and the single IMF variant (long dashed
    blue lines) produce reasonable fits to the observed LFs at both
    redshifts.  The \textit{insets} in each panel show the cumulative
    fraction of ionizing photons produced in galaxies brighter than,
    or fainter than, a given value of the $M_\mr{1500,AB}$ absolute
    AB magnitude (rising and falling curves, respectively). }
 \label{fig:LF_bursts}
\end{figure}

\begin{figure}
  \begin{center}
    \includegraphics[width=0.45\textwidth,keepaspectratio=true,     clip=true, trim=10 10 30 10]{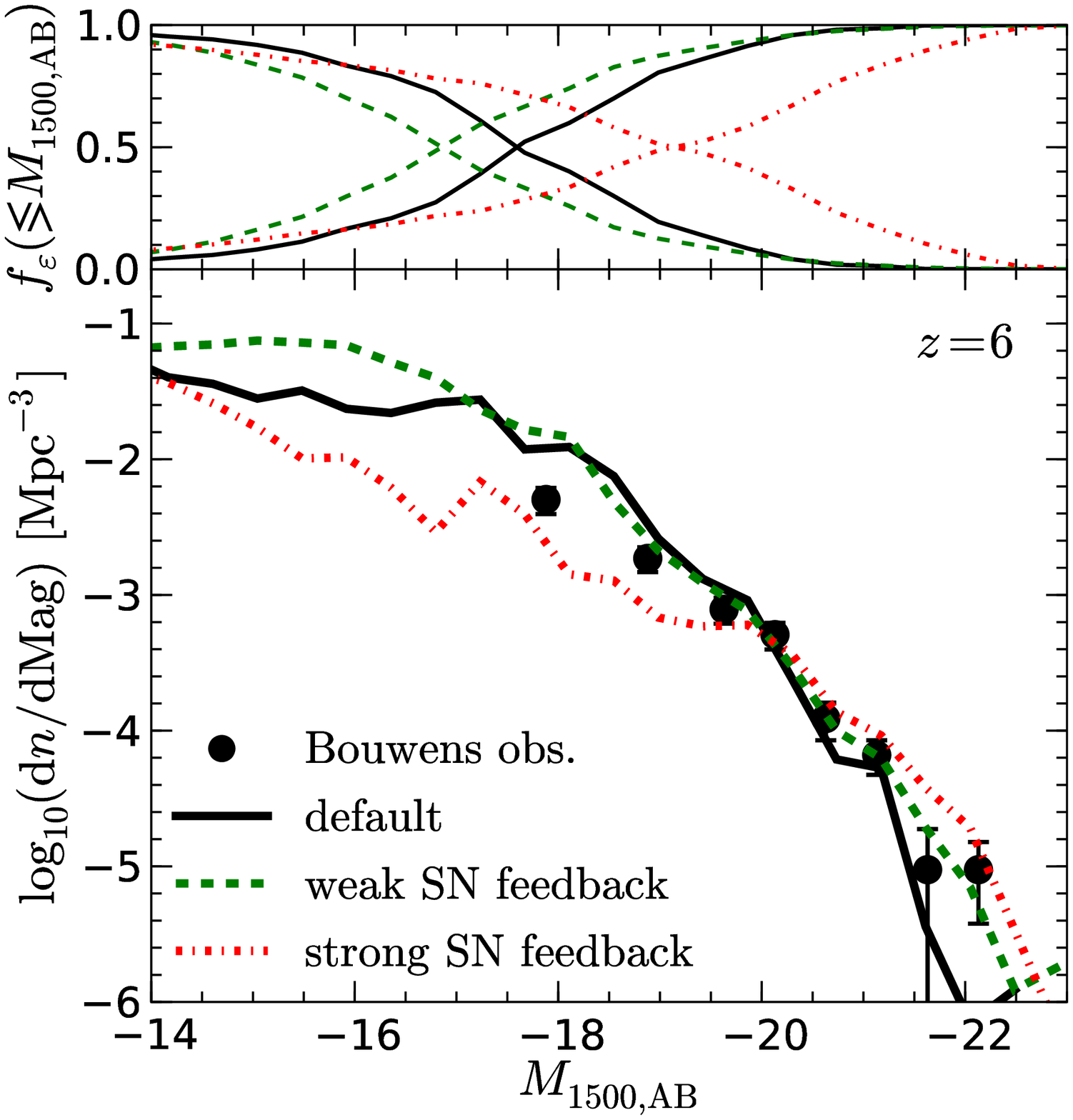}
    \includegraphics[width=0.45\textwidth,keepaspectratio=true, clip=true, trim=10 10 30 10]{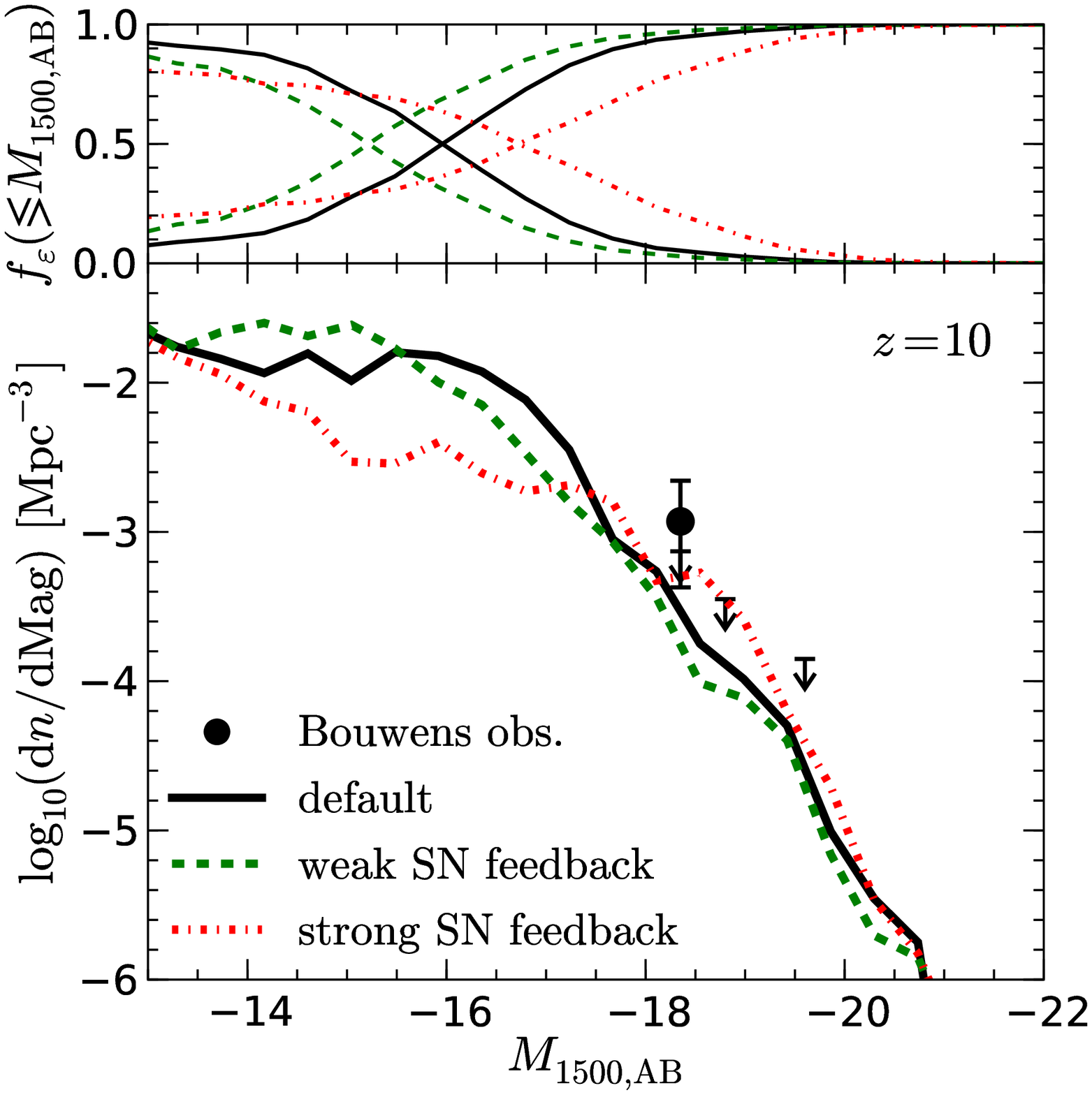}
  \end{center}
  \caption{ The effect of the supernova feedback parameters on the
    predicted rest-frame 1500\,\AA\ luminosity functions in the
    \Baugh\ model at redshifts 6 (top) and 10 (bottom), and models
    with weaker and stronger feedback (green and red lines,
    respectively); the corresponding emissivities were shown
    Fig. \ref{fig:SN_nphot}.  The data (solid points) are from Bouwens
    {\em et al.}, as in Fig.\ref{fig:LF_bursts}.  The weak feedback
    model (green dashed line) slightly over predicts the number of
    galaxies at $z\sim 6$, and the strong feedback model under
    predicts the numbers.  However at $z\sim 10$ the bright, observed
    end of the LF is equally well fit by all models.}
\label{fig:SN_LF}
\end{figure}

\begin{figure}
  \begin{center}
    \includegraphics[width=0.45\textwidth,keepaspectratio=true,     clip=true, trim=10 10 30 10]{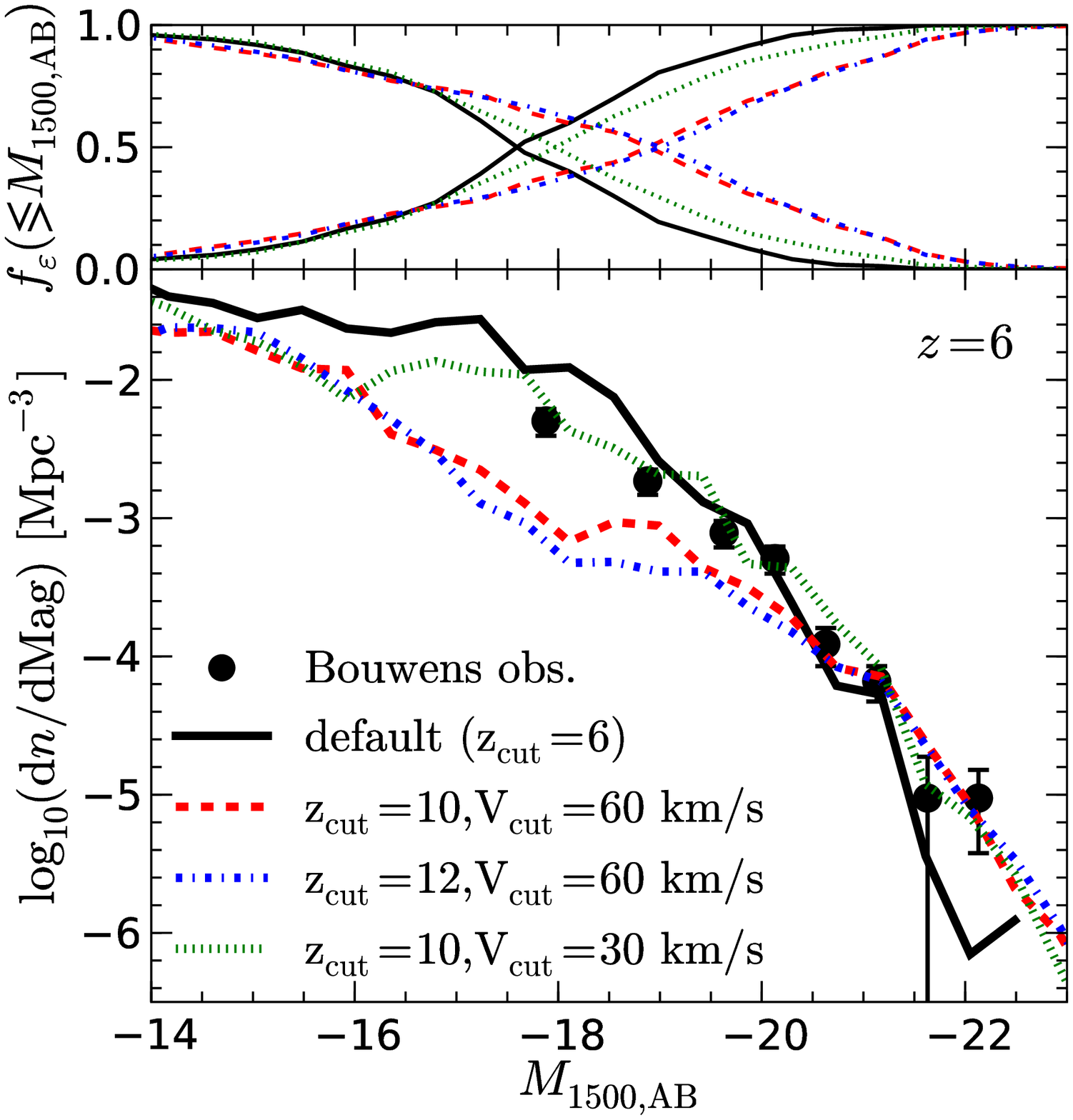}
  \end{center}
  \caption{The effect of photo-ionization on the predicted rest-frame
    1500\,\AA\ LF in the \Baugh\ model at redshift 6.  The models
    differ in their choice of reionization redshifts ($z_\mr{cut}$),
    and of the halo circular velocity below which galaxies are
    affected by photo-ionizing feedback ($V_\mr{cut}$). The
    corresponding emissivities were shown in
    Fig.\ref{fig:reion_params}.  The data (solid points) are from
    Bouwens {\em et al.}, as in Fig.\ref{fig:LF_bursts}. If galaxies
    with $V_\mr{cut}$=60~km~s$^{-1}$ are affected by suppression, then
    early reionization ($z_{\rm cut} \gtrsim 10$) can be ruled out by
    the current data, since then the predicted number density of
    galaxies at $M_\mr{1500,AB} \sim -18$ is $\sim 4$ times lower
    than observed (red lines). A more reasonable suppression scale of
    $V_\mr{cut}$=30~km~s$^{-1}$ is consistent with early reionization
    (green lines).}
\label{fig:zcut_LF}
\end{figure}

The Lyman-break colour-selection technique has proven to be very
effective for identifying large samples of star-forming galaxies at
high redshifts since its first application at $z\sim 3$
\citep{Steidel96}. This selection method was first applied at $z\sim
6$ by \citet{Bouwens03}, and recent deep near-IR imaging with {\em
Hubble Space Telescope} (HST) has been used to discover significant
numbers of candidate Lyman-break galaxies (LBGs) at $z\sim 7-8$, and a
few candidates at $z\sim 10$ \citep{Bouwens07, Bunker09a, Bouwens09a,
Bouwens09c, Oesch09}. We therefore now have direct detections of a
part of the galaxy population responsible for reionizing the
universe at $z\sim 6-10$. The companion paper by \citet{Lacey10b}
presents a detailed comparison of the predictions of \galform\ models
with observations of Lyman-break galaxies over the whole redshift
range $z=3-10$, including rest-frame far-UV luminosity functions,
sizes, masses and other properties. In this section, we investigate
what constraints can be put on the \galform\ parameters to which the
emissivity of ionizing photons $\epsilon$ is particularly sensitive
from observations of the rest-frame far-UV (1500\AA) luminosity
functions of $z\sim 6-10$ Lyman-break galaxies alone.  We also
investigate the extent to which the currently observed Lyman-break
galaxies contribute to the total emissivity of ionizing photons,
according to the \galform\ model.

\subsection{Effect of star formation  parameters and IMF}
The rest-frame 1500{\AA} broad-band \galform\ LFs at $z=6$ and $z=10$
are compared against the HST data on LBGs in
Fig.~\ref{fig:LF_bursts}. The default \Baugh\ model reproduces the LFs
at both redshifts, a considerable success. Clearly, starbursts are
crucial for bringing the 1500~\AA\ luminosities of the galaxies to the
observed levels (compare the red short dashed lines for the model
without bursts with the other two lines). These same bursts also
produce the bulk of the ionizing photons, as we showed in
Fig.~\ref{fig:bursts}.

Interestingly both the model with a top-heavy IMF in bursts (the
default model, black lines), and a model which uses the same
\cite{Kennicutt83} IMF in both quiescent galaxies and bursts (blue
dashed lines) fit the observed LFs nearly equally well at these
redshifts, notwithstanding the significant differences between these
models that we pointed out in, for example,
Fig.~\ref{fig:bursts_nphot}.  The reason for this is dust extinction:
the default model with the top heavy IMF produces more metals and
hence also more dust as compared to the \cite{Kennicutt83} IMF, and
the larger dust extinction partly compensates the larger intrinsic
far-UV luminosities \citep[see][for more
details]{Lacey10b}. Previously we found that a change in IMF affected the
ionizing emissivity considerably (Fig.~\ref{fig:bursts_nphot}), but
there we assumed that the escape fraction of ionizing photons $f_{\rm
{esc}}$, is simply a constant.  A physically motivated $f_{\rm {esc}}$
would presumably depend on galactic dust content, reducing the
difference between the top-heavy IMF and single IMF emissivities
\citep[see e.g.][]{Benson06} which would shift the completion of
reionization we found here to lower redshifts. We will examine these
issues in future work.

The currently detected candidate LBGs contribute only a small fraction
of the total emissivity of the whole population of galaxies predicted
by \galform\ at high-z. Even at $z \sim 6$ (top panel), galaxies
brighter than the current observational limit ($M_\mr{1500, AB, min}
\sim -18$) contribute only $\sim$ 40 per cent of the total ionizing
emissivity (solid black line in the top inset). If a single
\cite{Kennicutt83} IMF is assumed, that fraction is even lower
($\sim$ 20 per cent; long dashed blue line). At $z \sim 10$ (bottom
panel), more than 90 per cent of ionizing photons are emitted by
galaxies below the current detection limit for the default \Baugh\
parameters, and for a single IMF model that fraction is $\sim 95$ per
cent.

The \Baugh\ model predicts that the galaxies that produce the bulk of
the ionizing photons at $z\sim 10$ are intrinsically faint, with 50
per cent of ionizing photons produced in galaxies fainter than $m_{\rm
AB}\sim 31$ in the H-band.  Clearly it will be challenging to detect a
significant fraction of the galaxies that emit the photons that
reionized the Universe, even with the {\em James Webb Space
Telescope}, see {\em e.g.} the JWST white paper by Stiavelli {\em et
al.}\footnote{http://www.stsci.edu/jwst/science/whitepapers/}.

\subsection{Effect of supernova feedback parameters}
The strength of supernova feedback cannot be strongly constrained with
the current $z \gtrsim 6$ data \citep[Fig.\ref{fig:SN_LF}, see also][]{Lacey10b}. At the lowest redshift ($z=6$; top panel), the
faint end currently probed provides some constraints on the strength
of the supernova feedback, with the weak and strong models on either
side of the data. However, the $z=10$ data only probes the very
brightest galaxies, for which all three models predict very similar
LFs.

Of course, the supernova feedback parameters in \galform\ are strongly
constrained by even lower redshift data. However, the reader should
keep in mind that the emissivities we predict here are contingent on
the assumption that the basic physics of galaxy formation (in
particular the impact of supernova feedback on regulating star
formation) is the same at all redshifts. If for some reason this is
not true, the currently available observations at $z \gtrsim 6$ do not
probe sufficiently faint galaxies to determine the impact of
supernova feedback on the total emissivity produced by all galaxies.

\subsection{Effect of photo-ionization feedback parameters}
As discussed in Section~\ref{sect:photo}, the high-$z$ 1500 {\AA} LF
may hold information about the reionization history, if star formation
in galaxies is quenched once their surroundings are ionized. This is
illustrated in Fig.~\ref{fig:zcut_LF}. The $z=6$ LF is reasonably well
fit by the default \Baugh\ model, which assumes that reionization
occurs at $z_\mr{cut}=6$ (and hence for which there is no suppression in
Fig.~\ref{fig:zcut_LF}).

However, recent CMB measurements of the Thomson scattering optical
depth suggest reionization at $z \sim 10$, assuming an instantaneous
reionization model \citep{Komatsu10}. The \galform\ model with such
early reionization and $V_\mr{cut}=60$~km~s$^{-1}$ underpredicts the
faint end of the observed $z=6$ luminosity function by a considerable
amount, a factor $\sim 4$ for galaxies with $M_\mr{1500,AB}$ fainter
than -18. Clearly photo-ionization suppression is then too strong. But
we already argued that the default value of the halo circular velocity
below which galaxies are affected by photo-ionizing feedback
($V_\mr{cut}=60$~km~s$^{-1}$) is too high, with the hydrodynamical
simulations of \cite{Okamoto08} suggesting a much lower value of
$V_\mr{cut}=30$~km~s$^{-1}$. With this lower value of $V_\mr{cut}$,
the LF at $z=6$ is in good agreement with the data, even for an early
reionization redshift (green dotted line, see also \citet{Lacey10b});
in fact this model fits the $z=6$ data best. Noting that the CMB data
is the strongest current constraint on reionization, we argue that
this result gives an \textit{observational constraint} on the
characteristic strength of photo-ionization feedback that strengthens
the conclusion from current simulations.
 
The far-UV luminosity functions predicted by the \Baugh\ model and
presented here and in \cite{Lacey10b} show a very good agreement with
the $z \gtrsim 6$ data of \cite{Bouwens07, Bouwens08, Bouwens09a,
  Bouwens09b}. This is a significant success for a model for which
the parameters were chosen to match much lower redshift data, and
provides us with reasonable confidence in using the ionizing
luminosities predicted by this model in future, more detailed modeling
of the reionization process \citep{Raicevic10}.

We have seen that the \Baugh\ model predicts that the bulk of ionizing
photons is produced by galaxies significantly below the current
detection limit. It is a common practice to fit observed LFs with a
Schechter function, and use the fit to extrapolate the LF to fainter
galaxies. We show in the Appendix that this approach can lead to
significant errors in estimating the total emissivity, since the LFs
predicted by \galform\ deviate significantly from Schechter functions
in some ranges of luminosity, in particular due to the effects of
bursts. As a result, the Schechter fit parameters depend significantly
on the luminosity range over which the fit is done, and the total
emissivity estimated by extrapolating this fit is sensitive to the
minimum luminosity set by the observational detection limit.

\section{Conclusions}

We used the \citet{Baugh05} version of the \galform\ galaxy formation
model to compute the emissivity ($\epsilon$) of hydrogen-ionizing
photons in the redshift range relevant for reionization, $z \gtrsim
6$, and investigated the impact of changing some of the model
parameters from their default values.  A crucial element of this model
is that mergers between gas-rich galaxies increase $\epsilon$
dramatically compared to a model without bursts, mainly due to the
change to a top-heavy IMF in bursts assumed in the model. The
\citeauthor{Baugh05} model, with the same parameter values as used
here, has previously been shown to reproduce a wide range of observed
galaxy properties at lower redshifts.

The main points presented in the paper are:
\begin{itemize}

\item The \Baugh\ model produces enough ionizing photons to complete
  reionization by $ z \sim 10$ with galaxies alone, assuming a
  reasonable photon consumption (2 photons per hydrogen atom, allowing
  an average of 1 recombination per H atom) and a 20 per cent escape
  fraction of LC photons from galaxies (Fig.~\ref{fig:bau_bow_nphot}).

\item Starbursts are crucial for boosting the ionizing emissivity
    leading up to reionization. The majority of ionizing photons is
    produced in a relatively small fraction of galaxies at any given
    time that are bursting, and that are up to 5~dex brighter than
    non-bursting galaxies in halos of the same mass. Such bursts also
    increase the importance of intermediate-mass halos ($M \sim 10^9
    \msun$) compared to simpler models that do not include bursts
    (Fig.~\ref{fig:bursts_nphot}).

\item The top-heavy IMF used in the burst star formation mode is the
  main factor making the bursts so luminous, with $\sim 10$ times
  as many ionizing photons emitted per solar mass of stars formed as
  compared to the \cite{Kennicutt83} IMF. The change to a top-heavy
  IMF in starbursts was previously introduced in the model to
  reproduce the sub-mm galaxy counts at lower redshifts ($z\sim 1-3$),
  not the ionizing emissivity we discuss here, but it is crucial for
  completing reionization in agreement with current observational
  constraints.  The model with a single IMF reionizes $\Delta z
  \sim 2.5$ later than the default model
  (Fig.~\ref{fig:bursts_nphot}).

\item The assumed strength of supernova feedback has a strong impact
  on the ionizing emissivity, because the galaxies that dominate
  $\epsilon$ reside in relatively low-mass halos
  (Fig.~\ref{fig:SN_nphot}). This fact is of course well known at
  lower redshifts where strong feedback is required to reproduce the
  faint-end of the galaxy luminosity function \citep[e.g.][]{Cole00}, but is often ignored in reionization modeling, where a simple linear mass-luminosity relation is assumed.

\item As also shown in the companion paper by \citet{Lacey10b}, the
  \Baugh\ model reproduces the observed $z \sim 6-10$ rest-frame
  1500{\AA} luminosity functions well (Fig.~\ref{fig:LF_bursts}), with
  bursts a crucial ingredient in boosting the UV luminosities of
  galaxies to the observed levels.  The good agreement between the
  predicted and observed UV luminosity functions gives credence to
  using the model for computing $\epsilon$ as well. In the model,
  $\sim 90$ per cent of ionizing photons are produced by galaxies that
  are below the current HST detection limit at $z=10$, with 50 per
  cent of ionizing photons produced by galaxies fainter than $m_{\rm
  AB}\sim 31$ in the H-band. The intrinsic faintness of the sources
  will make it very challenging to detect a significant fraction of
  the galaxies that caused reionization, even with JWST.

\item The shape of the rest-frame far-UV luminosity function in the
  \Baugh\ model resembles a Schechter function, but with significant
  departures due to bursts. Given that the $z \gtrsim 6$ data only
  probe the bright end of this LF, extrapolating a Schechter function
  fit to estimate the contribution from galaxies below the detection
  limit can be inaccurate (see the Appendix).

\end{itemize}

As in all models of reionization, a significant uncertainty is the
fraction $f_{\rm {esc}}$ of ionizing photons produced by galaxies that
can actually escape into the IGM. We have intentionally used a simple
estimate for $f_{\rm {esc}}$, and our default value of 20 per cent is
somewhat higher than found observationally in lower redshift
observational studies \citep[e.g. $f_\mr{esc} \sim 10 \% $ for LBGs
at $z=3-4$][]{Steidel01}. A high dust content, one of the consequences
of using a top-heavy IMF, may decrease the escape fraction by as much
as an order of magnitude \citep{Benson06}. On the other hand, the
fraction of the ionizing photons that can escape into the IGM during a
burst could be significantly increased over the escape fraction during
quiescent star formation, due to the galactic wind driven by the
starburst.  Detailed numerical models that include turbulent motions
of gas in small galaxies find that $f_\mr{esc}$ can be as high as 0.5
- 1 {\em during} a burst \citep{Fujita03, Wise08, Wise09,
Razoumov10}. The enhancement of $f_{\rm esc}$ in bursts is likely to
be more dramatic for smaller galaxies than for larger ones, hence the
escape fraction is likely larger in small galaxies undergoing a
burst. If this is the case, then small, bursting galaxies will
dominate the Lyman-continuum emissivity even more. This strengthens
our main conclusion that small, starbursting galaxies can reionize
the Universe by $z\sim 10$. With this in mind, the value of
$f_\mr{esc} = 0.2$ that we used throughout this paper may even be
conservative.

As shown in the companion paper by \citet{Lacey10b}, the \Baugh\
\galform\ model reproduces the observed rest-frame 1500{\AA}
luminosity function of high redshift galaxies well over the whole
currently observed range $z=3-10$. A crucial ingredient in this model
is the boost in luminosity of galaxies as they undergo a minor or
major merger, when the stellar initial mass function becomes
top-heavy. This top-heavy IMF in bursts was originally introduced in
order to fit the counts of sub-mm galaxies at much lower $z\sim 2$,
and is necessary also to reproduce the observed high metallicity of
gas in $z\sim 0$ clusters of galaxies. A consequence is that bursts
generate the majority of Lyman-continuum photons. The model predicts
that starbursting galaxies with continuum UV magnitude $M_{1500, {\rm
AB}}\sim -16$, in halos of mass $\sim 10^9\,h^{-1}M_\odot$, dominate
the total emissivity at $z\sim 10$ (Fig. \ref{fig:bursts}). The
predicted properties of these galaxies have been analysed in more
detail in \cite{Lacey10b}. Those authors show that these galaxies have
stellar masses of $M_\star\sim 2\times 10^5\,h^{-1}M_\odot$, circular
velocities $V_c\sim 40$~km~s$^{-1}$, star formation rates $\dot
M_\star\sim 0.06\,h^{-1}M_\odot$~yr$^{-1}$, are gas dominated, $M_{\rm
gas}/M_{\rm baryon}\sim 1$, and have gas and stellar metallicities of
$\sim 4\times 10^{-3}$ and $\sim 3\times 10^{-3}$,
respectively. Assuming that on average approximately 2 ionizing photons are
required per hydrogen atom to reionize the Universe, a mean escape
fraction of 20 per cent is sufficient to reionize the Universe by
$z=10$.

\section*{Acknowledgments}
We would like to thank the \galform\ team for allowing us to run their
code, and for their constructive criticism. MR thanks John Helly,
Violeta Gonzalez-Perez, Wong Tam and Carton Baugh for practical help
and crucial discussions. During the work on this paper, MR was
supported by a grant from Microsoft Research Cambridge.

\bibliographystyle{mn2e}
\bibliography{paper2}

\section*{Appendix: Inferring ionizing emissivity from Schechter fits to the LF}

\begin{figure}
  \begin{center}
    \includegraphics[width=0.45\textwidth,keepaspectratio=true, clip=true, trim=0 10 10 10]{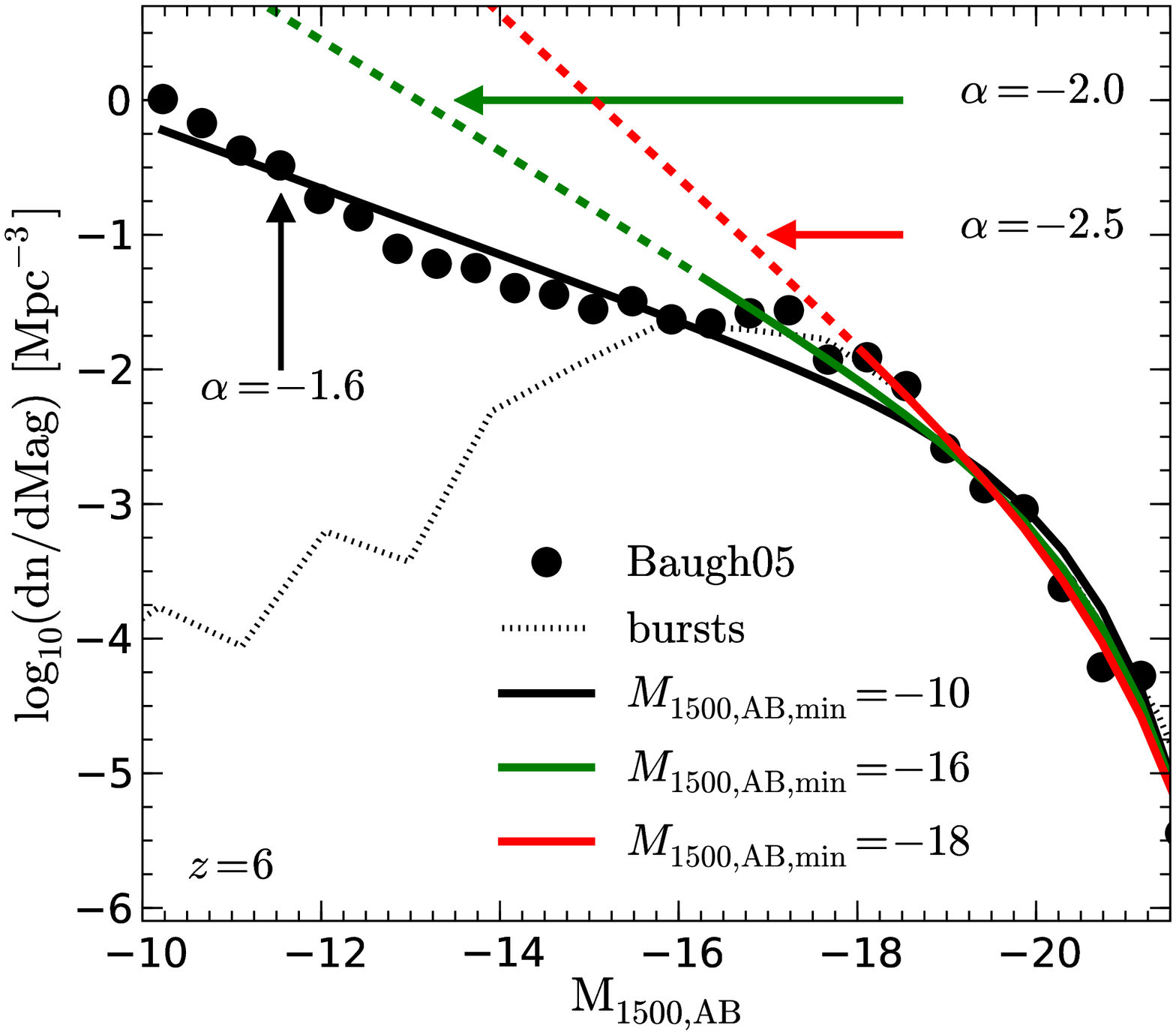}
  \end{center}
  \caption{Schechter function fits (coloured lines) to the far-UV LF
    in the \Baugh\ model (heavy dots) at $z=6$ over different absolute
    magnitude ranges, extrapolated to fainter luminosities (dashed
    lines).  Starbursts (dotted line) introduce a deviation in the
    shape of the LF from a Schechter function. Due to this feature,
    varying the minimum absolute magnitude employed in the Schechter
    fit results in very different estimates of the faint-end slope
    parameter, $\alpha^*$ (black, green and red lines correspond to
    minimum values of $M_\mr{1500,AB}$ of -10, -16 and -18 ,
    respectively). Extrapolation of the fits to fainter values can
    then lead to inaccurate estimates of luminosity density (see
    Fig.~\ref{fig:schechter_fit_params}), which in turn results in
    wrongly estimated LC emissivities. }
\label{fig:schechter_M_min}
\end{figure}

\begin{figure}
  \begin{center}
    \includegraphics[width=0.45\textwidth,keepaspectratio=true, clip=true, trim=0 10 10 10]{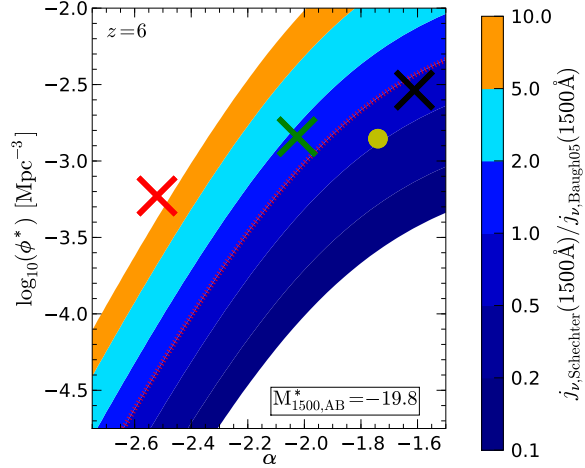}
  \end{center}
  \caption{Comparison of the 1500{\AA} luminosity densities,
  $j_{\nu}$, obtained by integrating the Schechter function fits to
  the \Baugh\ model LF at $z=6$, with the values obtained from
  integrating over the true model LF at the same redshift. The colour
  shading indicates the ratio of the luminosity density from the
  Schechter fit to the actual value in the model. In the Schechter
  fits, $M_{1500, AB}^*$ is held fixed at its best fitting value for
  the full luminosity range (black line in
  Fig.~\ref{fig:schechter_M_min}), while $\alpha$ and $\phi^*$ are the
  best fitting values for each choice of the minimum $M_\mr{1500,AB}$.
  Values of $(\alpha, \phi^*)$ along the {\em dotted red line}
  reproduce the actual luminosity density of the \Baugh\ model.
  Crosses mark the parameters of the three fits shown in
  Fig.~\ref{fig:schechter_M_min}, plotted in the same colors. When the
  whole luminosity range is used for the fit (black cross), the model
  luminosity density is reasonably well reproduced by the integral
  over the Schechter function fit. On the other hand, fits performed
  on a more limited luminosity range, as in
  Fig. \ref{fig:schechter_M_min}, lead to significant errors in the
  luminosity density estimate. The yellow dot shows the results
  obtained with the best fit parameters from \citet{Bouwens07} at
  this redshift.}
\label{fig:schechter_fit_params}
\end{figure}

In observational studies, the contribution of galaxies below the
detection threshold to the total ionizing emissivity is usually
estimated by fitting the observed LF with a Schechter function
\citep[e.g.][]{Bouwens07, Bunker09b}. At first glance, the far-UV LFs
predicted by the \Baugh\ model and shown in this paper are indeed
reasonably well represented by Schechter functions, as they have a
power-law shape at low luminosity, $\propto L^{\alpha}$, and an
exponential drop-off at the high luminosity end, $\propto
\exp(-L/L_\star)$. However, the LFs predicted by \galform\ are
\textit{not} in detail described well by Schechter functions. In
particular, in the \Baugh\ model, starbursts introduce a feature (a
\lq bump\rq) at $\sim 2$ magnitudes below $L_\star$ at high redshifts
(see \citet{Lacey10b} for more details). Due to this departure from
the Schechter shape, the result of fitting a Schechter function to a
\galform\ LF is strongly dependent on the luminosity range over which
the fit is done (Fig.~\ref{fig:schechter_M_min}). Assuming that
\Baugh\ is the \lq correct\rq\ model of the high-$z$ galaxy
population, the observational detection limits will then strongly
affect the predicted total ionizing emissivity, which relies on
extrapolating the contribution of the currently unobserved low
luminosity galaxies based on the faint-end slope $\alpha$ of the
Schechter fit. We want to investigate how much such extrapolations are
likely to be in error.

The estimated LC emissivity depends on more than just the LF shape,
with the choice of IMF and dust extinction being crucial yet only
weakly constrained by current observations. To focus only on the
uncertainty from the assumed LF shape, in
Fig. \ref{fig:schechter_fit_params} we show the dependence of the
1500{\AA} luminosity density, $j_{\nu}$, on the Schechter fit
parameters. All values of $j_{\nu}$ were obtained by integrating the
LFs over the magnitude range $-22 < M_\mr{1500,AB} < -10$. In this
figure, we vary only the normalization, $\phi^\star$, and the
faint-end power law slope, $\alpha$, and keep the characteristic
absolute magnitude, $M^\star_\mr{1500,AB}$, fixed, because the fits
shown in Fig.~\ref{fig:schechter_M_min} clearly have very similar
$M^\star_\mr{1500,AB}$ values.

With this procedure, a Schechter fit over the whole luminosity range
(down to $M_\mr{1500,AB} = -10$, black line in
Fig. \ref{fig:schechter_M_min}) of the \Baugh\ LF provides a good
estimate of the real luminosity density in the model (black cross in
Fig. ~\ref{fig:schechter_fit_params}; the luminosity density from the
fit is $\sim$ 20 per cent lower than the original model). If instead
the Schechter function is fit only to the brighter part of the LF
(green and red crosses, corresponding to $M_\mr{1500,AB,min}$ of -16
and -18, respectively), the faint-end slope of the model is strongly
overestimated. As a result, the luminosity density is also
overestimated in these cases, by factors $\sim$ 2 and 30 for
$M_\mr{1500,AB,min}=-16$ and -18 respectively. We note that the
\Baugh\ model predicts a total 1500{\AA} luminosity density at $z=6$ a
few times larger than estimates based on integrating the observed LF
only over the currently observed luminosity range
\citep[e.g.][]{Bouwens07}, but this difference shrinks if the observed
Schechter fits are extrapolated to lower luminosities (e.g. the yellow
circle in Fig.~\ref{fig:schechter_fit_params} shows the Schechter fit
found by \citeauthor{Bouwens07}, which implies a luminosity density
only 2 times lower than found in the model).

Some authors have concluded from integrating over the observed far-UV
LFs at $z \gtrsim 7$ that galaxies alone do not emit enough ionizing
photons to keep the universe ionized at these redshifts \citep[see
e.g.][]{Bunker09b}, but such conclusions seem premature, given that
they do not allow for galaxies fainter than the current detection
threshold or dust extinction or a different IMF slope.

This exercise aims to point out the danger of using Schechter function
fits to the observational data to estimate ionizing emissivity
produced by high-$z$ galaxies. The deviations of the LF from the
Schechter shape only add more uncertainty to the procedure which
already hinges on a number of unknowns, e.g. the choice of the IMF and
the dust extinction. This becomes even more important at higher
redshifts, where the LF is even more poorly constrained by current
observational data.

\label{lastpage}

\end{document}